\documentclass[pdftex,twocolumn,epjc3]{svjour3}          

\RequirePackage[T1]{fontenc}

\smartqed  

\RequirePackage{graphicx}
\RequirePackage{mathptmx}      
\RequirePackage{flushend}
\RequirePackage[numbers,sort&compress]{natbib}
\RequirePackage[colorlinks,citecolor=blue,urlcolor=blue,linkcolor=blue]{hyperref}

\journalname{Eur. Phys. J. C}

\usepackage{amssymb,amsmath}
\usepackage{slashed,color}
\usepackage{bbold}
\usepackage{subfigure}
\usepackage{multirow}

\usepackage{dcolumn}
\usepackage{bm}
\usepackage{slashed} 
\usepackage{epsfig}
\usepackage{verbatim}
\usepackage[utf8]{inputenc}
\usepackage[T1]{fontenc}
\usepackage[normalem]{ulem}

\newcommand{\beq}{\begin{eqnarray}}
\newcommand{\eeq}{\end{eqnarray}}

\newcommand{\bmp}{\noindent\begin{minipage}{16cm}}
\newcommand{\emp}{\end{minipage}\vskip 7mm} 

\newcommand{\GeV}{\mbox{ ${\mathrm{GeV}}$}}

\newcommand{\ifb}{\mbox{ ${\mathrm{fb^{-1}}}$}}

\newcommand{\Tr}{\mbox{Tr}\;}
\newcommand{\be}{\begin{eqnarray}}
\newcommand{\ee}{\end{eqnarray}}
\newcommand{\gt}{\widetilde{g}}

\newcommand{\st}{s_\theta}
\newcommand{\ct}{c_\theta}
\newcommand{\stt}{s_{2\theta}}
\newcommand{\ctt}{c_{2\theta}}
\newcommand{\SU}{\mbox{SU}}
\newcommand{\SP}{\mbox{Sp}}
\newcommand{\SO}{\mbox{SO}}
\newcommand{\UU}{\mbox{U}}

\newcommand{\ii}{\mathrm{i}}

\def\eq#1{{eq.~(\ref{#1})}}
\def\eqs#1#2{{eqs.~(\ref{#1})--(\ref{#2})}}
\def\fig#1{{fig.~(\ref{#1})}}

\hypersetup{draft} 
\begin{document}
\title{Sigma-assisted low scale composite Goldstone-Higgs}

\author{Diogo Buarque Franzosi\thanksref{e1,addr1}
        \and
       Giacomo Cacciapaglia \thanksref{e2,addr2}
                \and
        Aldo~Deandrea\thanksref{e3,addr2}
}
\thankstext{e1}{buarque@chalmers.se}
\thankstext{e2}{g.cacciapaglia@ipnl.in2p3.fr}
\thankstext{e3}{deandrea@ipnl.in2p3.fr}

\institute{Department of Physics, Chalmers University of Technology, 
	Fysikg\aa rden, 41296 G\"oteborg, Sweden \label{addr1}
          \and
         Universit\'{e} de  Lyon, Univ. Claude Bernard Lyon 1, CNRS/IN2P3, IP2I Lyon, UMR 5822, F-69622, Villeurbanne, France
\label{addr2}
}

\date{Received: date / Accepted: date}

\maketitle


\begin{abstract}

We show that the presence of a lightish scalar resonance, $\sigma$, that mixes with the composite Goldstone-Higgs boson can relax the typical bounds found in this class of models. This mechanism, inbred in models with a walking dynamics above the condensation scale, allows for a low compositeness scale $f \gtrsim 400$~GeV, corresponding to a 
misalignment angle $\st \lesssim 0.6$, contrary to the common lore of a smaller angle. 
According to recent lattice results, the light $\sigma$ emerges thanks to a near-conformal phase above the condensation scale, consistent to the requirements from flavour physics.
We study this effect in a general way, showing that it appears in all cosets emerging from an underlying gauge-fermion dynamics, in the presence of top partial compositeness.
The scenario is testable both on the Lattice and experimentally, as it requires the presence of a second broad Higgs-like resonance, below 1 TeV, that can be revealed at the LHC in the $ZZ$ and $t\bar{t}$ channels.

\end{abstract}

\maketitle

\section{Introduction}

Modern composite Goldstone-Higgs (CH) models~\cite{Bellazzini:2014yua,Panico:2015jxa} are promising candidates to dynamically and naturally generate the electroweak (EW) symmetry breaking: a condensate breaks, non perturbatively, a global symmetry of the strong sector that includes the EW gauge symmetry. 
The misalignment of the vacuum condensate with respect to the EW group thus induces a hierarchy between the compositeness scale $f$ and the EW scale $v \approx 246$~GeV,  parameterised as 
\begin{equation}
v=f \st\,,  
\end{equation}
where $\theta$ is the misalignment angle~\cite{Dugan:1984hq}, and we adopt the short-hand notation $\sin \theta = \st$, $\cos \theta = \ct$ and $\tan \theta = t_\theta$. 
The Higgs boson $h$ appears as a pseudo-Nambu-Goldstone boson (pNGB), thus explaining its lightness compared to the other composite states and its approximate EW doublet nature~\cite{Kaplan:1983fs}. 
In comparison, in Technicolor models~\cite{Weinberg:1975gm,Susskind:1978ms,Dimopoulos:1979es}, which are matched in the limit $\theta \to \pi/2$, the role of the Higgs can only be played by a light singlet scalar resonance~\cite{DiVecchia:1980xq,Dietrich:2005jn,Belyaev:2013ida} or a dilaton-like light state~\cite{Yamawaki:1985zg,Bando:1986bg,Dzhikiya:1986kk}.

One of the main model-building challenges
encountered in generic CH models written at the effective Lagrangian level is obtaining the ``little hierarchy'' $v \ll f$,
in compliance with EW precision observables (EWPOs). Due to large corrections to the oblique $S$ parameter~\cite{Peskin:1990zt,Peskin:1991sw,Barbieri:2004qk}, the compositeness scale needs to be sizeably larger than the EW scale, yielding a fairly model-independent bound $\st\lesssim 0.2$~\cite{Barbieri:2012tu,Grojean:2013qca,Arbey:2015exa}. 
This, however, requires a tuning in the parameters of the model, which can happen in the top sector alone~\cite{Matsedonskyi:2012ym} or by tuning the current mass term of the underlying fermions~\cite{Galloway:2010bp,Cacciapaglia:2014uja}. We remark that the pNGB Higgs mass is always of order $m_h \approx f \st = v$, and its precise value is encoded in a generally incalculable strong form factor. 
A lot of effort has been dedicated to this issue in the literature, with many mechanisms designed to minimise the fine tuning in the CH potential (for recent works, see refs.~\cite{Csaki:2017cep,Csaki:2017jby}).

In this work we take an orthogonal approach, and show that a mild hierarchy $v \lesssim f$
may be an intrinsic property of most CH models that feature a nearly conformal (or walking) phase right above the condensation scale. As we will see, the key player is a light scalar resonance in the spectrum.  Being an intrinsic property of the strong dynamics, the mass and couplings of this state are completely determined by the underlying dynamics, thus they cannot be tuned to be in the favourable parameter region. 
Only Lattice results or other non-perturbative techniques will allow to determine if such parameters are in the right ballpark.
We do not try to estimate the fine tuning associated to this mechanism, judging that solving a several orders of magnitude hierarchy is worthy the price of a fine-tuning, while we address the more phenomenological question of how low the compositeness scale could be while complying with all experimental tests.

The manuscript is organized in the following way. In Sec. 2 we discuss the vacuum alignment and argue that the natural value of the misalignment angle would be in tension with experimental data if not for the presence of a light scalar state mixing with the Higgs boson. In Sec. 3 we give a prescription to describe the pNGB and the light scalar. In Sec. 4 we discuss the constraints on the model including Higgs measurements and EWPO. In Sec. 5 we study the possibility to observe the light scalar decaying into $ZZ$ or $t\bar{t}$ at the LHC. In Sec. 6 we provide an interpretation of the required parameters in terms of a light scalar originated from confining dynamics.
We finally offer our conclusions in Sec. 7.

\section{Misalignement and dynamics}

Typically, the top quark couplings to the strong sector dominate the misalignment dynamics at low energies. 
If the top mass is generated by contact interactions {\it \`a la} extended Technicolor~\cite{Dimopoulos:134187}, then the natural alignment is towards the Technicolor vacuum $\st = 1$. It has been shown in ref.~\cite{Arbey:2015exa} that the introduction of a lightish scalar resonance $\sigma$, that mixes with the pNGB Higgs, can alleviate the tension between EWPOs and the Technicolor vacuum.  Increasing evidence of the presence of a light scalar state in theories with an infra-red conformal phase are being collected on the lattice~\cite{Hasenfratz:2016gut,Athenodorou:2016ndx,Aoki:2017fnr,Appelquist:2018yqe}, and by the use of gravitational duals~\cite{Elander:2017cle,Elander:2017hyr} and of holographic models~\cite{BitaghsirFadafan:2018efw}. Such state, which may or may not be a dilaton~\footnote{A dilaton is a pNGB associated to the spontaneous breaking of conformal invariance at quantum level.}, necessarily mixes with the Higgs boson. Furthermore, a conformal phase, also called ``walking''~\cite{Holdom:1981rm}, can help alleviating the flavour issue of CH models~\cite{Matsedonskyi:2014iha,Cacciapaglia:2015dsa} by increasing the gap between the compositeness scale and the scales of flavour violation. Finally, partial compositeness~\cite{Kaplan:1991dc} has been identified as a promising mechanism to give a large mass to the top quark provided that the fermion operators that linearly mix with the top feature a large anomalous dimension in the walking window. 

All the features listed above, therefore, point towards realistic CH models where a light $\sigma$ is always present and it should be included as part of the minimal set-up. Note that here {\it light} refers to mass scales around or below $f$, thus between the EW scale $v$ and roughly $1$~TeV. In this letter we analyse the possibility of having low scale compositeness
in CH models with top partners thanks to the presence of such a light $\sigma$. 
The details of the composite Higgs potential generated by the top couplings are very model dependent, as the results vary greatly depending on the top partner representation and on the coset. In the following, to demonstrate how the mechanism work, we will consider a simplified scenario based on some reasonable assumptions.

Firstly, we will consider the three minimal cosets deriving from a gauge-fermion underling description, even though we will see that the coset structure does not play a crucial role in our discussion.
Secondly, we will consider cases where the top mass depends on the misalignment angle as follows:
\begin{equation}
m_t \propto f s_{2\theta}\,.
\end{equation}
This case occurs in many instances, and we refer the reader to refs.~\cite{Golterman:2017vdj,Alanne:2018wtp,Agugliaro:2018vsu} for a survey of different top partner representations in the cosets $\SU(4)/\SP(4)$ and $\SU(5)/\SO(5)$.
Thirdly, we will assume that the potential is dominated by top loops, thus being proportional to $m_t^2$:
the natural minimum is, therefore, at $\theta=\pi/4$ ($\st = 1/\sqrt{2}$) and not at the Technicolor limit.
Besides the issue with EWPOs, a large $\st$ also induces large modification to the pNGB Higgs couplings to SM states. Interestingly, such corrections for the $W$ and $Z$ couplings are universal and model-independent~\cite{Liu:2018vel}: this is due to the fact that they are determined by the $\theta$--dependence of the masses. The reduced couplings to massive gauge bosons $V = W^\pm, Z$ and top (normalised to the SM values) are, therefore, equal to 
\begin{equation}
\kappa_V= \frac{\partial_\theta v}{v} = \ct\,, \qquad \kappa_t = \frac{v}{f m_t} \partial_\theta m_t =\frac{\ctt}{\ct}.
\label{eq:hcoup}
\end{equation}
Note that the coupling to the top vanishes at $\theta = \pi/4$. This property can be easily understood: the minimum of the potential is given by $0 = \partial_\theta V(\theta) \propto m_t (\theta)\ \partial_\theta m_t (\theta)$, thus at the minimum one has either $m_t (\theta) = 0$ or $\partial_\theta m_t (\theta)=0$. As $\kappa_t \propto \partial_\theta m_t$, a vacuum with non-zero top mass implies $\kappa_t = 0$. 
The recent detection of the $t\bar{t} h$ production channel by CMS~\cite{Sirunyan:2018hoz} and ATLAS~\cite{Aaboud:2018urx} that definitely proves $\kappa_t \neq 0$,  therefore, rules out the most natural minimum. \footnote{The indirect probe from the top-loop induced gluon coupling is not effective as the strong sector can give additional contributions.}
The presence of a light $\sigma$ that mixes with the pNGB Higgs can alleviate both issues of EWPOs and the Higgs coupling modifications. 

\section{Modelling the strong sector}

In order to show the effect of the light $\sigma$, we will use an effective field theory approach to describe the properties of the light pNGB degrees of freedom, plus the $\sigma$, without relying too much on the details of the strong sector.

The minimal chiral Lagrangian describing the pNGBs and the $\sigma$ is given, schematically, by~\cite{Cacciapaglia:2014uja,Hansen:2016fri}
\begin{eqnarray}
{\cal L}&=& k_G(\sigma) \frac{f^2}{8} D_\mu \Sigma^\dagger D^\mu \Sigma - \frac{1}{2}(\partial_\mu\sigma)^2-V_M (\sigma)\nonumber\\
&+&k_t(\sigma)\ \frac{y_L y_R f C_y}{4\pi}(Q_\alpha t^c)^\dagger \, \Tr[(P_Q^\alpha\Sigma^\dagger P_t \Sigma^\dagger)]+\text{h.c.}\nonumber \\
&-& k_t^2 (\sigma) V_{t} - k_G^2 (\sigma) V_{g} 
\,, \label{eq:chiFT}
\end{eqnarray}
where $\Sigma = e^{i \Pi/f}\cdot \Sigma_0$ is the linearly transforming pNGB matrix defined around the vacuum $\Sigma_0$. The term in the second line is responsible for the top mass with the spurions $P_Q$ and $P_t$ ($\alpha$ is an $SU(2)_L$ index). $V_M$ is a potential for $\sigma$, and $V_{t,g}$ are the terms in the pNGB potential generated by the top and gauge loops respectively.
We do not require $\sigma$ to be a dilaton, even though our general expression can accommodate light dilaton effective Lagrangians~\cite{Golterman:2016lsd,Appelquist:2017wcg}. 
The effective Lagrangian is essentially the same for the different cosets. In the simplest coset $\SU(4)/\SP(4)$ the pNGB matrix contains the would-be Higgs $h$, a pseudo-scalar singlet $\eta$ and the eaten NGBs $\pi^a$, 
while additional pNGBs are present in larger cosets.

We work in a basis where none of the fields defined in \eq{eq:chiFT} are allowed to develop a non-zero vacuum expectation value, i.e. both pNGBs and $\sigma$ are defined around the proper vacuum. In particular, $\Sigma_0$ contains the misalignment along the Higgs direction that breaks the EW symmetry, and the $\sigma$-potential $V_M(\sigma)$ includes a tadpole that balances up the contribution of the pNGB potential terms. Furthermore, we normalise the $\sigma$ coupling functions such that $k_i (0) = 1$ and define $k'_i = \left. f \frac{\partial k_i}{\partial \sigma} \right|_{\sigma = 0}$, $k''_i = \left. f^2 \frac{\partial^2 k_i}{\partial \sigma^2} \right|_{\sigma = 0}$, etc.

We choose the top spurions in \eq{eq:chiFT} so that the top mass reads
\begin{equation}
m_t = \epsilon_{tL} \epsilon_{tR} \frac{C_y f \st c_\theta}{4 \pi}\,,
\label{eq:mt}
\end{equation}
where $\epsilon_{tL/R}$ encode the degree of compositeness of the left- and right-handed tops respectively and $C_y$ is a strong dynamics form factor.
This expression is common to all the specific scenarios we consider.
The pNGB potential reads~\cite{Alanne:2018wtp}
\begin{equation}
V_t = -  C_t' f^4 \st^2 \ct^2 + \dots\,, \qquad
V_g = -C_t' \delta f^4 \ct^2 + \dots\,. 
\end{equation}
where, for later convenience, we have defined
\begin{equation}
\delta \equiv \frac{C_g (3g^2+g'^2)}{2 C'_t}\,, \qquad C_t'\equiv \frac{C_t \epsilon_{tR}^2 \epsilon_{tL}^2}{(4\pi)^2}\,, \label{eq:deltadef}
\end{equation}
where $C_g$ and $C_t$ are strong dynamics form factors.~\footnote{$C_g$ has been calculated on the lattice for a specific underlying model in ref.~\cite{Ayyar:2019exp}.}
This leads to the minimum condition
\begin{equation}
\frac{\partial V}{\partial \theta} = 2 f^3 \st \left(f C'_t (c_{2 \theta} -\delta) \ct \right)=0\,.
\label{eq:min-theta}
\end{equation} 
The zero at $\st = 0$ would imply that the EW symmetry is unbroken ($\theta=0$). However, as we expect the top loop to drive the potential to break the EW symmetry (i.e., $C_t >0$), the minimum of the potential sits at 
\begin{equation} \label{eq:minimum}
\delta=\ctt\,,
\end{equation}
as long as $0 < \delta < 1$. It is reasonable to assume that the top loops dominate, i.e. $\delta \ll 1$, thus the most likely minimum should sit at $c_{2\theta} \sim 0$, which is therefore the most ``natural'' misalignment in this class of models.
Moving away from it would require either to enhance $\delta$ by suppressing the top contribution to the potential, or by adding a sizeable current mass $m_\psi$~\cite{Galloway:2010bp,Cacciapaglia:2014uja}, thus falling into the fine tuning issue of CH models\footnote{The effect of fermion current mass on the vacuum alignment is further discussed in \ref{sec:currentmass}.}.
We will show that there exist allowed regions in the parameter space where the minimum can stay close to the most natural value, $\theta = \pi/4$.

From \eq{eq:chiFT} it is straightforward to compute the masses and couplings of the pNGB Higgs and the singlet by taking derivatives with respect to $\theta$ and $\sigma$. A mixing between the two is always present, proportional to $k'_t$ and $k'_G$. 
We find the following relation between the mixing term and the mass eigenvalues $m_{h_{1/2}}$:
\begin{equation} \label{eq:deltaA}
\frac{k_G'-k_t'}{t_{2\theta}} \equiv \delta_A \frac{m_{h_2}^2-m_{h_1}^2}{2m_{h_1}m_{h_2}}\,, 
\end{equation}
with $|\delta_A|\leq 1$. See~\ref{app:scalardetail} for more details about the scalar mass mixing and the $\delta_A$ parameter.
The couplings of the two mass eigenstates to the gauge bosons $V=W,Z$ and top quarks read
\begin{eqnarray}
\kappa^{h_1}_V&=&\ct c_\alpha + (k_G'/2) \st s_\alpha\,, \label{eq:kVh1}\\
\kappa^{h_1}_t&=& \frac{\ctt}{\ct}c_\alpha +   k_t' \st s_\alpha\,, \label{eq:kth1}
\end{eqnarray}
where $\alpha$ is the mixing angle between the two states $h$--$\sigma$ and the mass eigenstates. 
The couplings of the heavier $h_2$ are obtained by $\alpha \to \pi/2 + \alpha$.
The absolute value of the angle $\alpha$ is small, bounded by $|\tan(2\alpha)|\leq \frac{2m_{h_1}}{m_{h_2}}$ (see \fig{fig:tan2a} in \ref{app:scalardetail}).
A derivative-coupling of $\sigma$ to the pNGBs is also present, leading to an effective coupling
\begin{eqnarray}
g_{\sigma \pi^2}&=& k_G' \frac{p_1\cdot p_2}{v} \st  + {\cal O}\left(\frac{m_{\pi}^2}{v}\right)\,, 
\label{eq:sigmapi}
\end{eqnarray}
where $\pi$ includes the Higgs and other pNGBs, which will be crucial for a correct calculation of the width of the heavier state. All the formulas above are universal and valid for all CH cosets, where only the total width of $h_2$ will depend on the total number of pNGB in the cosets (assuming they are all lighter).

\begin{table}[tb]
\begin{tabular}{c|c|c|}
 & Constraint & Value \\
\hline
\multirow{3}{*}{i - } & Perturbativity & $|k'_{G,t}| < 4 \pi $ \\
                            & Unitarity~\cite{BuarqueFranzosi:2017prc} & $\gamma \equiv \frac{m_{h_2}}{4\sqrt{\pi} f} \lesssim 1$ \\
                            & Small width & $\Gamma_{h_2} < m_{h_2}$ \\
\hline
\multirow{3}{*}{ii -} & Combined fit (Run-I)~\cite{Khachatryan:2016vau} &   $\kappa_V^{h_1}=1.035\pm 0.095$ \\
                           & $t\bar{t}h$ production~\cite{Sirunyan:2018hoz,Aaboud:2018urx} &  $\kappa_t^{h_1} = 1.12^{+0.14}_{-0.12}$ \\
                           & Higgs width~\cite{Khachatryan:2016vau} & $\mathcal{BR} (h_1 \to \mbox{BSM}) < 0.32$ \\
\hline
\multirow{3}{*}{iii -} & \multirow{3}{*}{EWPOs~\cite{Haller:2018nnx}} & $S = 0.04 \pm 0.08$ \\
                                             & & $T = 0.08 \pm 0.08$ \\
                                             & & correlation $\sigma_{TS} = 0.92$ \\
\hline
\end{tabular}
\caption{List of theoretical and experimental constraints on the model.} \label{tab:constraints}
\end{table}

\section{Constraints on the model}
\begin{figure*}
\centering
\begin{tabular}{c c c}
\includegraphics[width=0.4\textwidth]{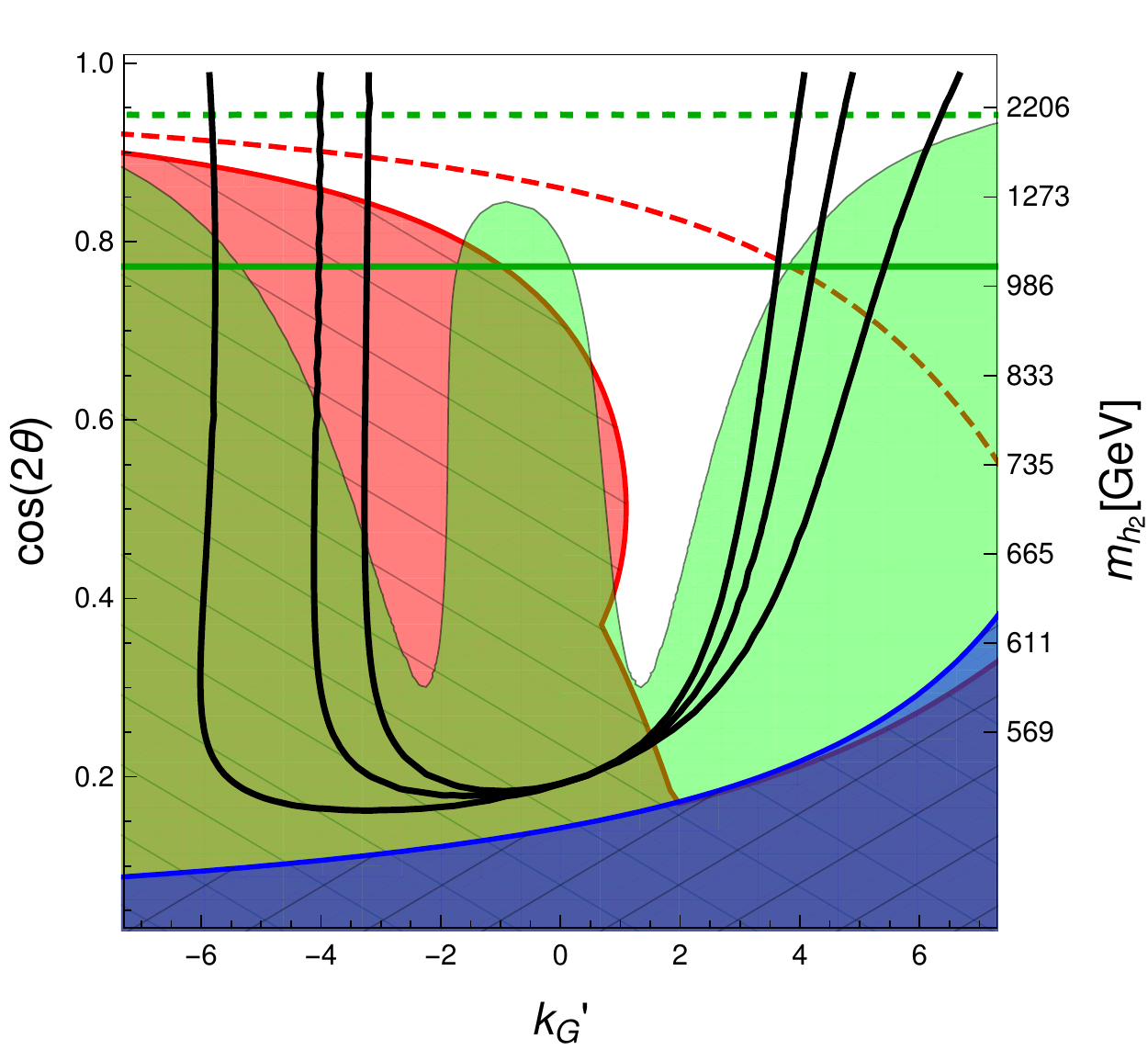} & \phantom{xxxxx} &
\includegraphics[width=0.4\textwidth]{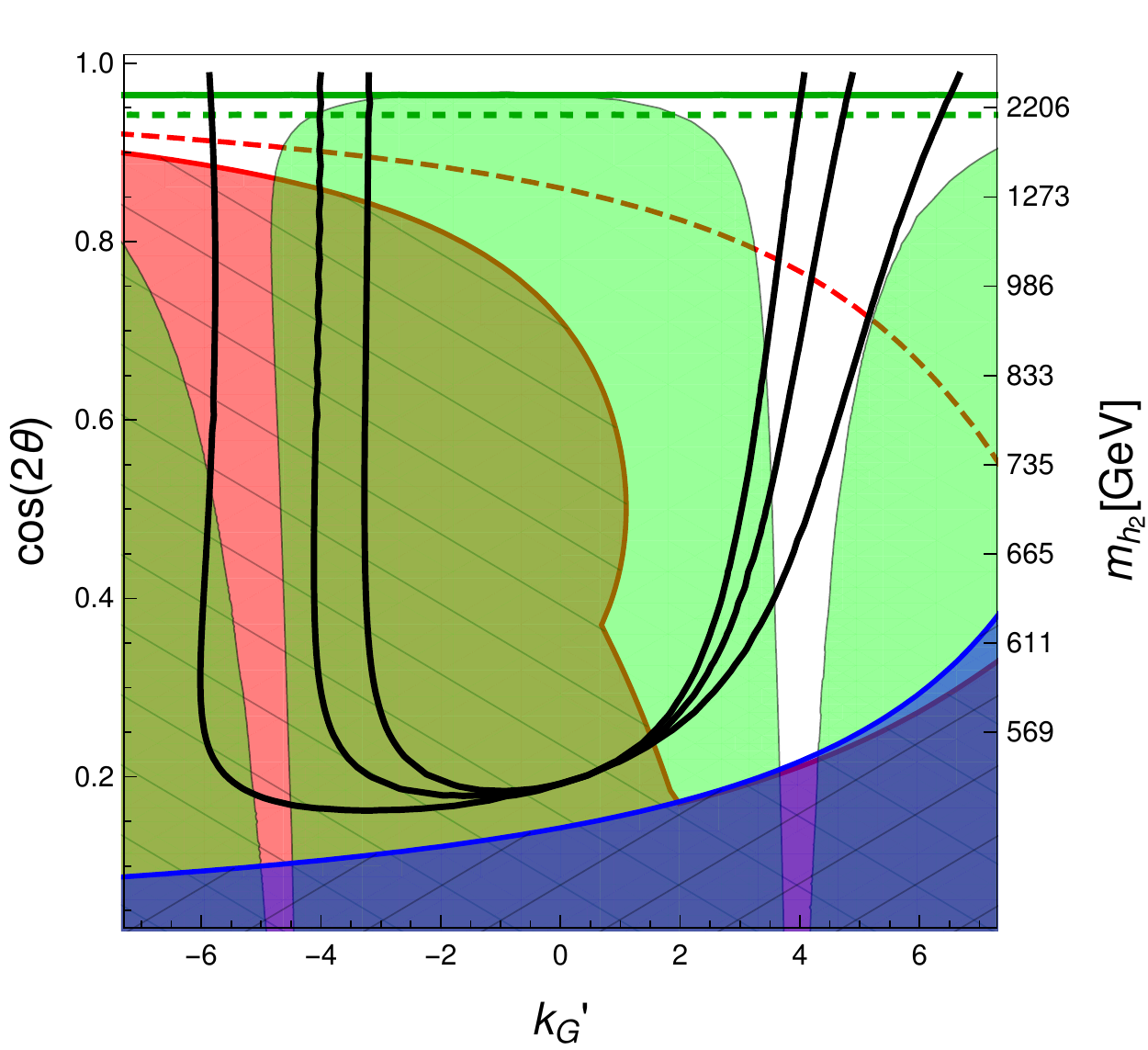}\\
(a) & &  (b) 
\end{tabular}
\caption{Excluded regions in $k'_G$--$\cos (2\theta)$ for a benchmark model. Plot (a) has negative $\Delta S_\rho$ ($r=1.1$), while (b) has positive $\Delta S_\rho$ ($r=0.9$). The black curves indicate where $\Gamma_{h_2}=m_{h_2}$ for the 3 minimal cosets. The excluded regions come from perturbativity (blue), Higgs couplings (red) and EWPOs (green). The green dashed (continuous) lines show the EWPO bounds if the $\sigma$ ($\sigma$ plus vectors) is removed from the model.
In both plots $\delta_A=-0.9$, $\gt=3$, and $\gamma = 0.2$. 
}
\label{fig:constraints}
\end{figure*}

We now discuss the constraints on the general model presented in the previous section. The scalar sector $h_{1,2}$ has only 4 free parameters: the mass $m_{h_2}$ (we fix $m_{h_1} = 125$~GeV), the misalignment angle $s_\theta = v/f$, and the two $\sigma$ couplings, $k'_G$ and $k'_t$.
We probe the parameter space of the model by imposing the constraints listed in Table~\ref{tab:constraints}.
The main goal is to determine whether large values of the misalignment angle $\theta$ are allowed.  We recall that the mass of $\sigma$ and its couplings are not free parameters, but fully determined by the underlying dynamics. Thus our goal is to find the favourable values, which can be tested on the Lattice in specific cases.
The excluded regions in $\cos(2\theta)\times k_G'$ plane are shown in \fig{fig:constraints} with further discussion in the text below.

\subsection{Perturbativity and unitarity}

We require that all couplings stay in the perturbative regime. To this effect, we demand that all the $\sigma$-couplings respect $|k_i'|<4\pi$, $i=t,G$.
Furthermore,
to guarantee perturbative unitarity we demand that the pNGB scattering remains perturbative up to the condensation scale $4 \pi f$. 
This requirement is connected to the one above, as we expect that the $\sigma$ plays a crucial role in taming the growth with energy of the amplitude, like in QCD~\cite{Harada:1995dc}.
We will base our estimate on the leading order calculation, even though radiative corrections typically tend to increase the amplitude and push the resonance mass to lower values~\cite{BuarqueFranzosi:2017prc}.
Neglecting effects from the potential, which are irrelevant at high energies, the asymptotic behaviour of the pNGB scattering in the sigma channel (projection on zero isospin and angular momentum, $I=J=0$) is given by
\begin{eqnarray}
a^{(0)}_{A0}(s) \approx \frac{N_f}{32\pi f^2}\ s\,,
\label{eq:aI0}
\end{eqnarray}
where $N_f$ is the number of Dirac fermions in the underlying gauge-fermion theory. We thus require that the mass of the $\sigma$ lies below the energy scale where the above amplitude grows larger than $1$.  This bound can be expressed as 
\begin{equation} \label{eq:gamma}
\gamma\equiv \frac{m_{h_2}}{4\sqrt{\pi} f}\leq \sqrt{\frac{2}{N_f}}\,,
\end{equation}
where $\sqrt{\frac{2}{N_f}} = 1$, for the minimal case $N_f=2$.

Finally, we also require that the heavy scalar width remains small compared to the mass. Although a broad state is a plausible scenario, we require $\Gamma/M<1$ in order to trust our perturbative calculations.
In fact, for heavy masses ($m_{h_2}\approx m_\sigma \gg v$) the total width of the heavier scalar $h_2$ is dominated by the $\sigma$ component, and given by the model-independent derivative coupling in \eq{eq:sigmapi}.
It reads
\begin{equation}
\frac{\Gamma_{h_2}}{m_{h_2}} \approx k_G'^2\ \frac{N_\pi m_{h_2}^2}{128\pi f^2} + k_t'^2\ \frac{3 m_t^2}{8\pi f^2}\,,
\end{equation}
with $N_\pi$ being the number of light pNGBs.
The first term can be recognised as the partial width into pNGBs, while the second is the partial width into tops. For our perturbative treatment of the mixing between $h$ and $\sigma$, we need to make sure that the width remains small or comparable to the mass of the scalar. 

\subsection{Higgs measurements}

Since the discovery of the Higgs boson, both ATLAS and CMS have been measuring its couplings with increasing precision. These measurements provide relevant limits on any model that modifies the Higgs sector.
The reduced couplings  of $h_1$ to $V=W^\pm, Z$ and the top are given in \eqs{eq:kVh1}{eq:kth1}.
In general, the couplings to light fermions, like the bottom and tau, will also be affected, while direct contributions of the strong sector may affect the couplings to gluons and photons, which are loop induced in the SM. However, the details are model dependent. Here we want to be conservative, so we will extract only bounds on $\kappa_V$ and $\kappa_t$ that are independent on other measurements.  

For the coupling to vectors, we use the combined fit of ATLAS and CMS after Run-I~\cite{Khachatryan:2016vau} and extract the bound on $\kappa_V$ from the most general fit. The coupling to tops is bounded indirectly by the measurement of the gluon fusion cross section. However, if we allow for a generic contribution to the gluon coupling from new physics, the only solid bound comes from the observation of the $t\bar{t}h$ production mode~\cite{Sirunyan:2018hoz,Aaboud:2018urx}. We thus translate the most stringent bound on the signal strength, $\mu=1.26^{+0.31}_{-0.26}$ from CMS, to a bound on $\kappa_t$. The Higgs coupling bound we impose are indicated in Table \ref{tab:constraints}. 
Note that Run-II bounds on the couplings to vectors are becoming more constraining, however a combination is still not available and we refrain to do it as taking into account systematics of the experiments cannot be reliably done.

\begin{figure}
\centering
\includegraphics[width=0.4\textwidth]{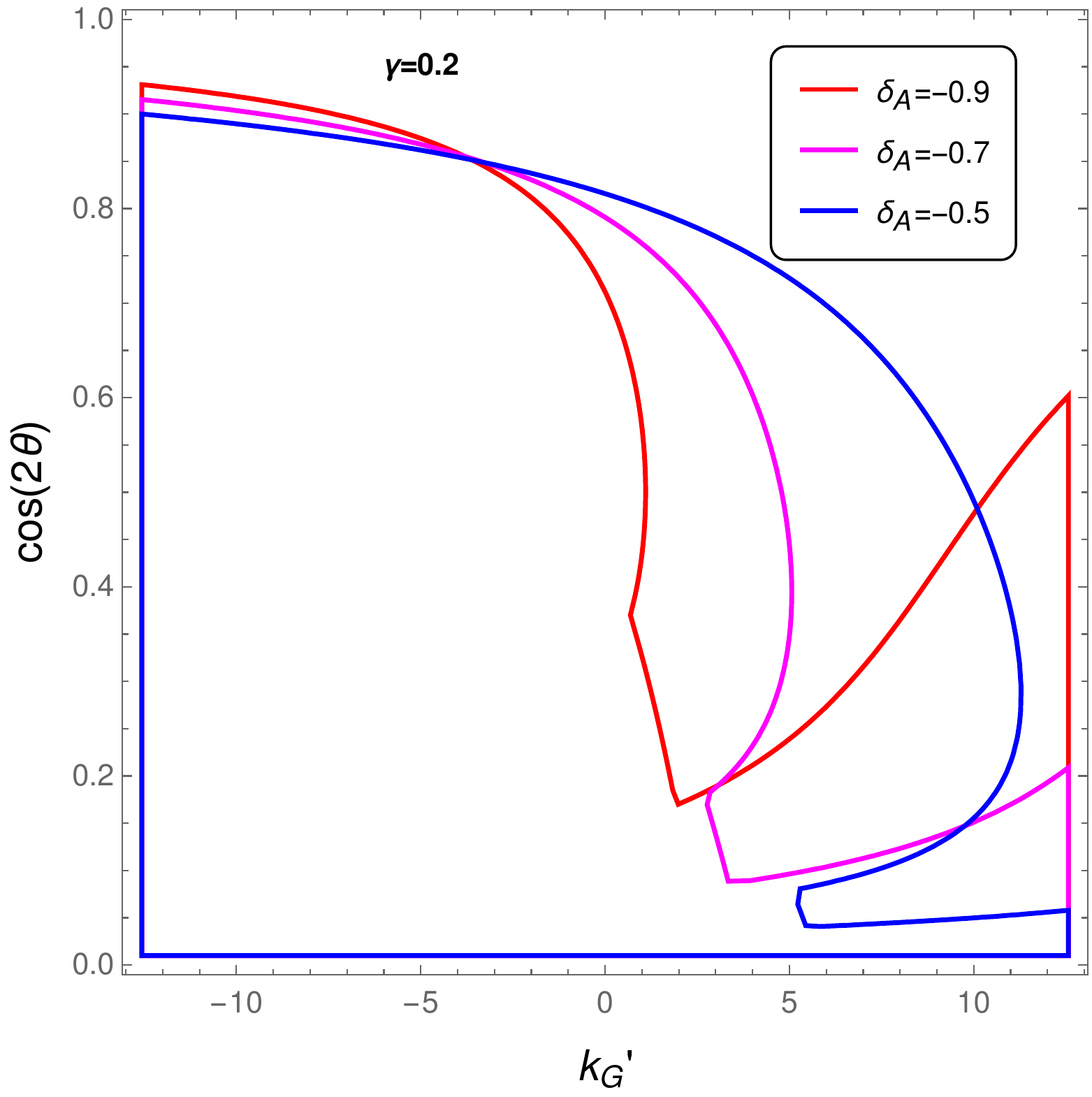}
\caption{Higgs coupling excluded regions for different values of $\delta_A$ and $\gamma=0.2$. 
}
\label{fig:higgscoup}
\end{figure}

The constraints at $2\sigma$ level on the parameter space coming from the Higgs coupling measurements are shown in red in \fig{fig:constraints}, where the recent observation of $t\bar{t}h$ production mode plays a crucial role in constraining the top Yukawa.~\footnote{$\kappa_t^{h_1}$ is indirectly constrained from the global fit~\cite{Khachatryan:2016vau}. As the uncertainties are similar to the ones from the direct measurements, we do not expect stricter bounds and therefore conservatively rely on the direct bound.}
 This constraint is the most sensitive to the mixing parameter $\delta_A$. In \fig{fig:higgscoup} we test different values of this parameter and display the excluded regions as coloured contours. It can be noticed that a value $|\delta_A|\gtrsim 0.7$ is required for the mechanism to work (note that changing the sign $\delta_A\to -\delta_A$, the exclusion plot is left-right flipped requiring positive $k_G'$), mainly due to the Higgs-top coupling.

The constraint from the new physics decays of the Higgs can be relevant if additional pNGBs are lighter than $m_{h_1}/2$: the potential impact of this constraint is indicated with a dashed red line in panels (a) and (b) of \fig{fig:constraints} in the case of a very light $\eta$ for $\SU(4)/\SP(4)$, note however that this bound is eluded by adding a small current mass $m_\psi$ that can make $\eta$ heavier without sizeably affecting the misalignment (see \ref{sec:currentmass} for more details).

\subsection{Electroweak precision observables}
The effect of the Higgs coupling modification and $h_2$--loops can be described in the oblique formalism as~\cite{Arbey:2015exa}
\begin{eqnarray}
\Delta S&=& \frac{1-(\kappa_V^{h_1})^2}{6\pi}\ \log\frac{\Lambda}{m_{h_1}} - \frac{(\kappa_V^{h_2})^2}{6\pi}\ \log\frac{\Lambda}{m_{h_2}} + \Delta S_{\rm TC}\,, \nonumber
\\
\Delta T&=& \frac{-3 (1-(\kappa_V^{h_1})^2)}{8\pi c^2_{\theta_W}}\ \log\frac{\Lambda}{m_{h_1}} + \frac{3 (\kappa_V^{h_2})^2}{ 8\pi c^2_{\theta_W}}\ \log\frac{\Lambda}{m_{h_2}}  \,,
\end{eqnarray}
where $\Delta S_{\rm TC} = N_D \st^2/(6\pi)$ is the contribution of the strong sector~\cite{Peskin:1991sw,Sannino:2010ca}, with the $N_D$ factor counting the number of EW doublets in the underlying theory 
and we use $\Lambda^2=2\pi^2 f^2$ as the compositeness scale.

Vector and axial-vector resonances are also known to contribute to the oblique parameters and cause cancellations \cite{Casalbuoni:1995yb,Contino:2015mha}.
Following ref.~\cite{Franzosi:2016aoo}, we computed the contribution to the $S$-parameter  under the assumption of Vector Meson Dominance  for the three cosets $\SU(4)/\SP(4)$, $\SU(5)/\SO(5)$ and $\SU(4)\times \SU(4)/\SU(4)$.  We find that the result is parametrically the same, and it amounts to replacing the term $\Delta S_{\rm TC}$ with
\begin{eqnarray}
\Delta S_{\rm TC} \to \Delta S_\rho&=& \frac{16\pi(1-r^2)\st^2}{2(g^2+\gt^2)-g^2(1-r^2)\st^2}\,,
\label{eq:rhoewpo}
\end{eqnarray}
with $\gt$ and $r$ being non-perturbative parameters of the chiral Lagrangian (see \ref{app:EWPO} for details on the calculation). A cancellation thus happens for $r>1$, where the new correction to $S$ is negative.  Note that this result is not present elsewhere in the literature.

Other pNGBs also contribute to the EWPOs at one loop level. For example in the coset $\SU(4)\times \SU(4)/\SU(4)$ they contribute~\cite{Ma:2015gra}
\begin{eqnarray}
\Delta S_\pi &=& -\frac{\st^2}{4\pi}\,, \\
\Delta T_\pi &=& \frac{\st}{16\pi s_W^2}\left(\frac{m_{H^\pm}^2-m_A^2}{m_W^2}\right)\log\left(\frac{\Lambda^2}{m_\pi^2}\right)\,,
\end{eqnarray}
with $H^\pm$ and $A$ being the charged and pseudo-scalar component of the second pNGB Higgs doublet generated by the coset. 
This effect is sub-dominant compared to the Higgs and the vector and axial-vector ones.
In $\SU(5)/\SO(5)$ extra care must be taken to avoid that the EW triplet pNGB gets a vacuum expectation value thus generating a large tree-level contribution to the T parameter~\cite{Agugliaro:2018vsu}.

The exclusion from EWPOs is shown by the green region in \fig{fig:constraints}.
These plots reveal the presence of two ``valleys'' reaching large $\theta$, one of which compatible with other bounds. This is \emph{genuinely due to the scalar} $\sigma$, as proven by comparing the two panels in \fig{fig:constraints}, which differ by the sign of $\Delta S_\rho$ that is negative in (a) and positive in (b). Thus, the effect of the vector cancellation for $\Delta S_\rho<0$ is merely to shift the valleys towards smaller $k'_G$. For comparison, we show with dashed green lines the bounds without $\sigma$, and in solid green without both $\sigma$ and the vectors.

In \fig{fig:constraints} we also show the values of the $h_2$ mass corresponding to $\cos (2\theta)$: in order to achieve 
large misalignment, a sub-TeV scalar resonance is required that may be accessible at the LHC. {\it This is to be considered a prediction of sigma-assisted low-scale CH models.}
This state will dominantly decay into two massive gauge bosons, $W^+ W^-$ and $ZZ$, and $t\bar{t}$, however current LHC searches cannot directly apply because of the large width. 
The $ZZ$ resonance CMS search~\cite{Sirunyan:2018qlb} explores widths up to 30\% of the mass and could cover this region, however larger widths would need to be included in the search. We also estimate that resonant production of $t\bar t$ could be competitive to the $ZZ$ channel for larger couplings to gluons. 
The presence of this sub-TeV broad resonance in $ZZ$ and $t\bar{t}$ is a smoking gun of this scenario, and our estimates clearly motivate dedicated large-width searches at the LHC.
In the next section we estimate the present reach of the LHC in observing this state using the available searches and tools.

\section{Direct searches for the heavy scalar at the LHC}
\label{sec:heavyscalar}

The main signature of sigma-assisted low-scale CH models is the presence of a second heavier ``Higgs'' $h_2$, which may be observed at the LHC. Its mass is a free parameter, however, as we have seen, it is limited below a TeV by requiring perturbative control of the effective theory. The production mechanisms are the same as for the SM Higgs, namely gluon fusion (ggF) and vector boson fusion (VBF),  with associated production with tops to a lesser extent.

The production of $h_2$ via ggF is difficult to estimate due to its loop-nature - besides the top loop generated by the top coupling in
\eq{eq:kth1}, 
the strong dynamics can give an additional direct coupling term.  To understand the structure of the latter, we analyse the possibility that it is generated dominantly by a loop of a heavy top-like resonance $T$, a.k.a. top partner. The coupling of the $\sigma$ will have the form
\begin{equation} \label{eq:kappaT}
k_T(\sigma)\ M_T T\bar{T} = \left( M_T + k'_T \frac{M_T}{f} \sigma + \dots \right) T\bar{T}\,,
\end{equation}
where the top partner mass $M_T = g_T f$, with 
$g_T = {\cal O}(1) < 4 \pi$.  
Thus, one can define a reduced coupling 
\begin{equation}
\kappa_T^\sigma = k'_T g_T \st \,,
\end{equation}
which is explicitly suppressed by a power of the misalignment angle. This suppression will appear in the effective coupling to gluons, which will thus be $\kappa_g^{h_2} \propto \st$.

To estimate the production cross section of $h_2$ we use the N$^3$LO result for ggF~\cite{Anastasiou:2016hlm} and the NNLO for VBF production~\cite{Bolzoni:2010xr,Bolzoni:2011cu}, and rescale the SM Higgs production cross section as follows:
\begin{equation}
\sigma=\sigma_0^{gg} \frac{|\kappa_t^{h_2} A_F(\tau_t)+\kappa_g^{h_2}|^2}{|A_F(\tau_t)|^2} + \sigma_0^{VBF}(\kappa_V^{h_2})^2\,,
\end{equation}
where $A_F(\tau_t)$ is the standard loop amplitude for the top quark in the SM.
Following the argument above, in the following we will fix the strong dynamics contribution to the coupling to gluon as $\kappa_g^{h_2}/\st =$ const.

\begin{figure*}
\centering
\includegraphics[width=0.32\textwidth]{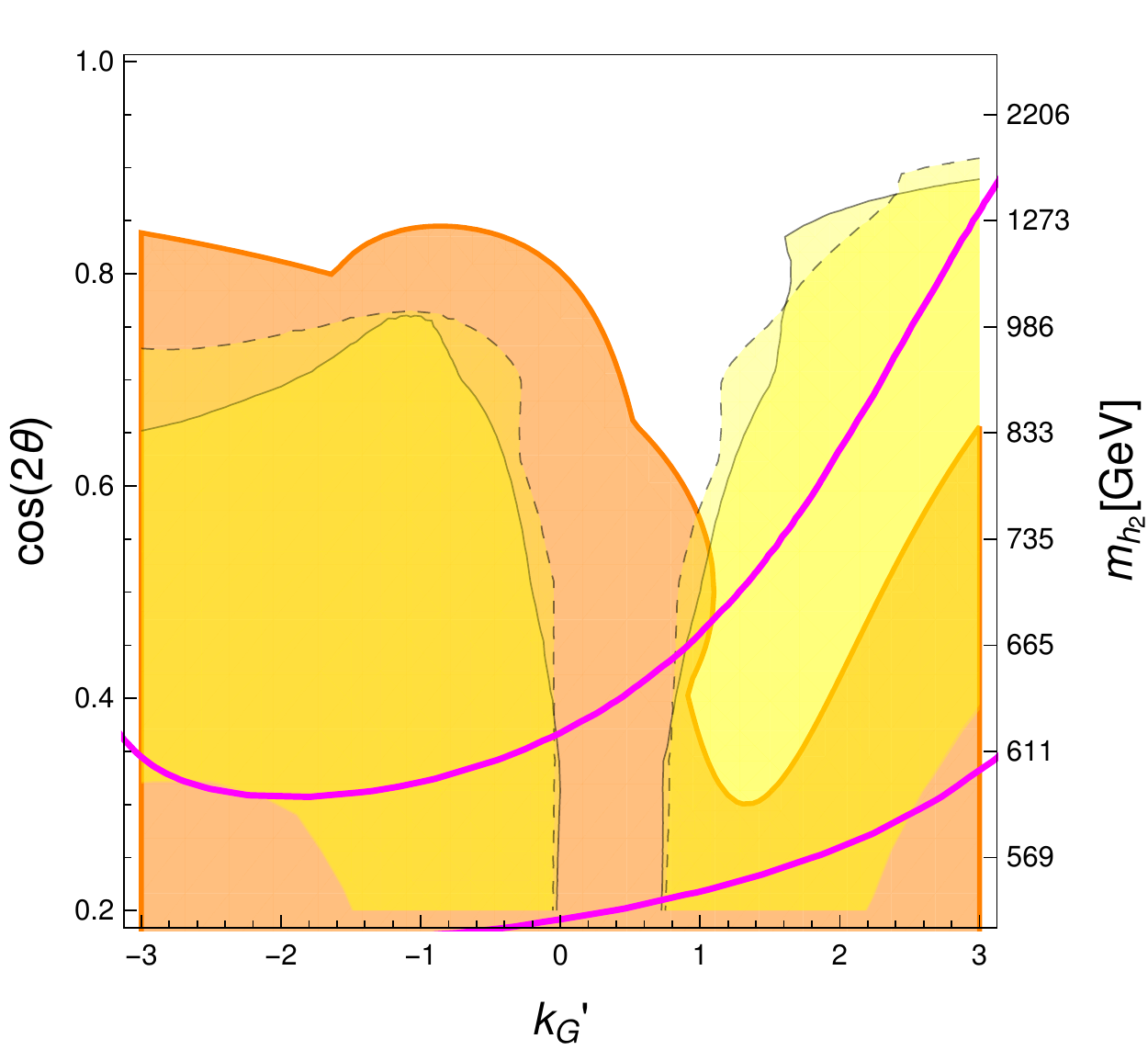}
\includegraphics[width=0.32\textwidth]{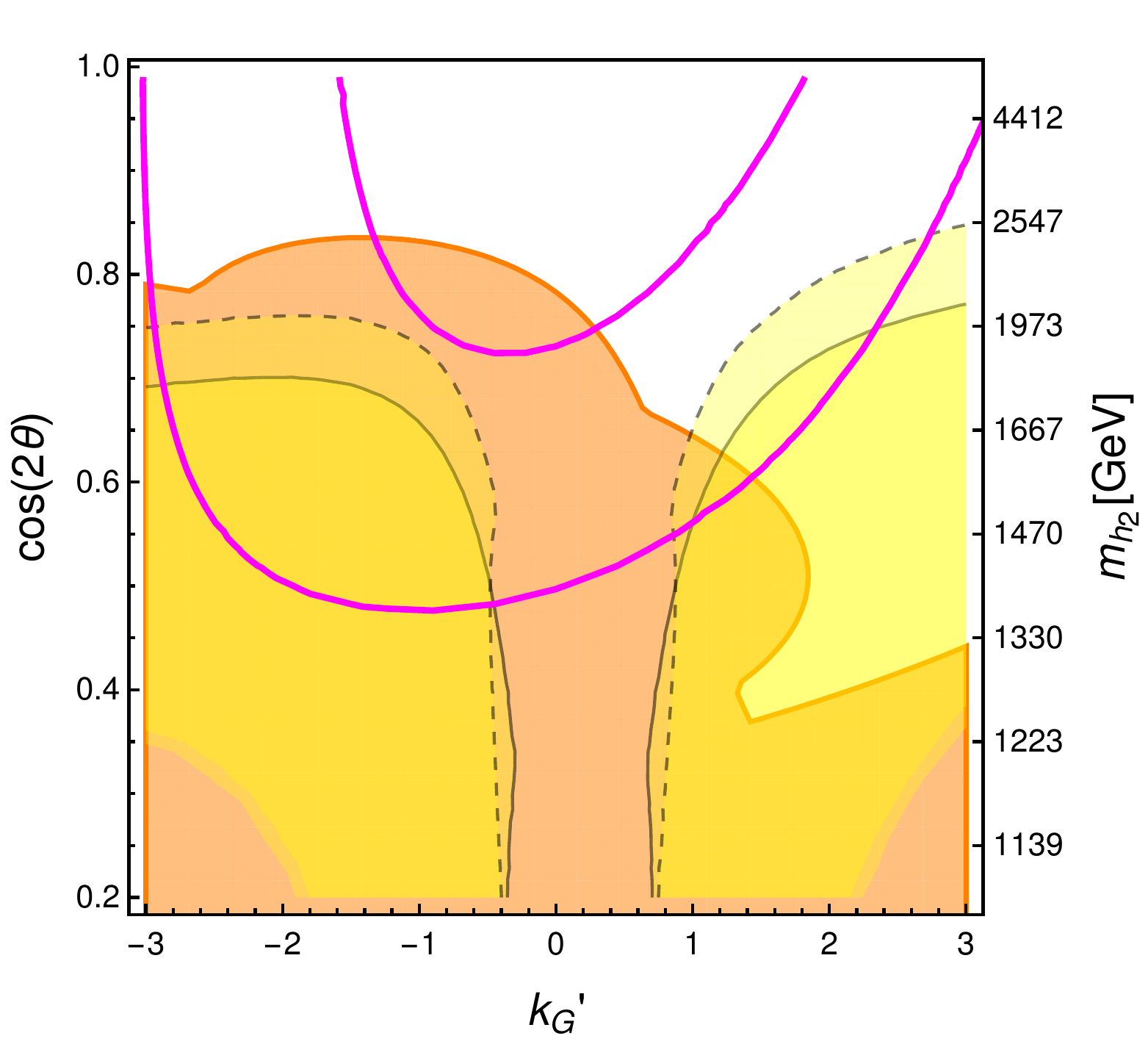}
\includegraphics[width=0.32\textwidth]{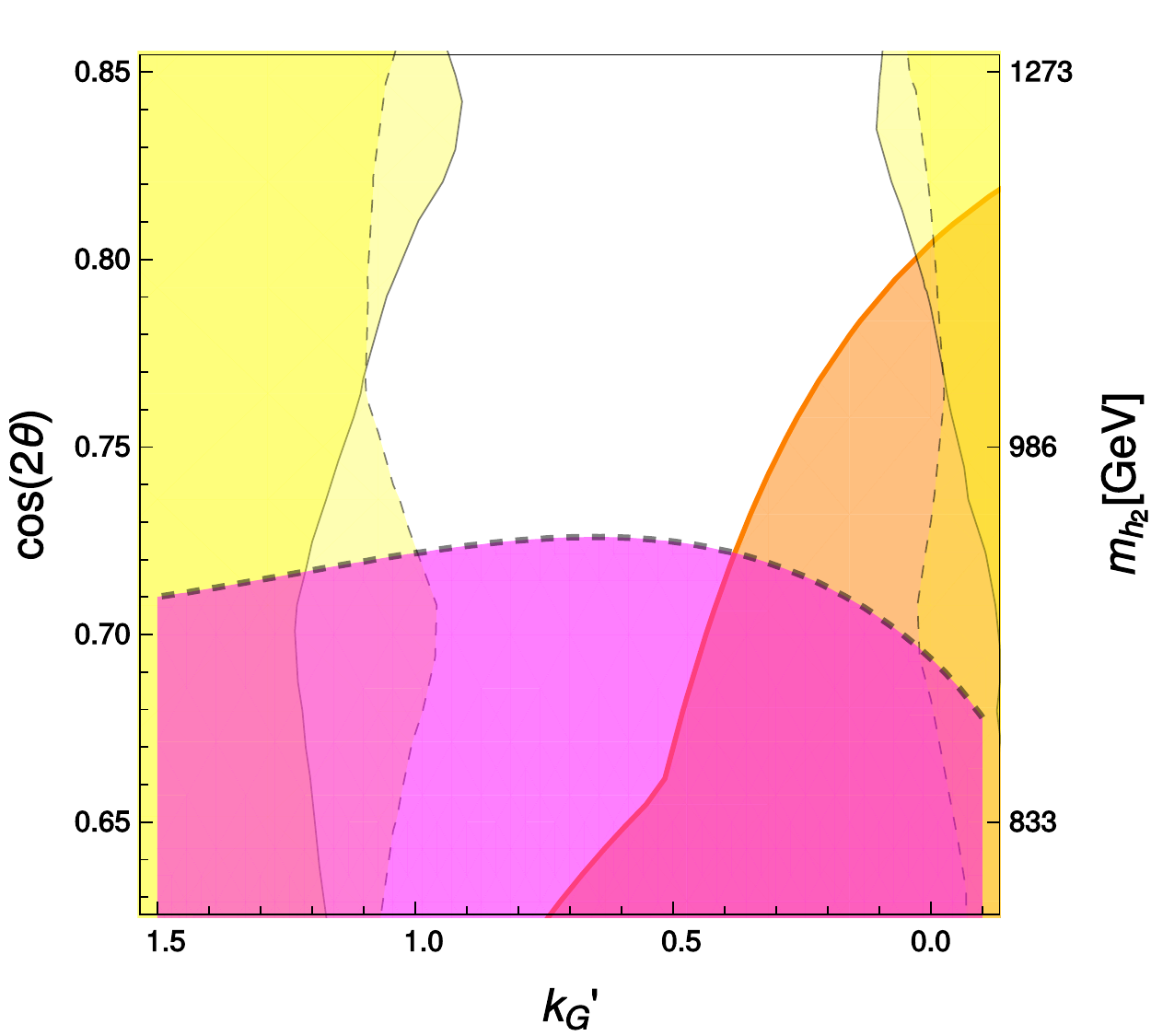}
\caption{Region potentially excluded at 95\% CL by the CMS $ZZ$ search, in yellow,  superimposed on $t\bar{t}$ search, in magenta in the right plot, and previous indirect bounds in orange. The yellow shaded region with solid contour correspond to the limit for $\Gamma/M=0.3$, while the dashed line corresponds to $\Gamma/M=0$. 
We used $r=1.1$, $\gt=3$, $\delta_A=-0.9$ and in the \emph{left panel} $\gamma=0.2$, $\kappa_g=\st$, in the \emph{middle panel} $\gamma=0.4$, $\kappa_g=\st$, and in the \emph{right panel} $\gamma=0.2$, $\kappa_g=5\st$ (note the different range and inverted order in $k_G'$). }
\label{fig:excl-summary}
\end{figure*}

The strongest experimental constraint on a heavy Higgs-like resonance comes from $ZZ$ searches. The main issue with reinterpretation of the experimental result is due to the fact that $h_2$ in this class of models tends to be very broad. In ref.~\cite{Sirunyan:2018qlb} the CMS collaboration considered broad scalar resonances decaying into $ZZ$ final states, with widths up to $\Gamma/M < 0.3$. 
We can thus extract expected exclusions by simply comparing the production rates of the $ZZ$ final states directly with the experimental results. This is shown in \fig{fig:excl-summary} for the parameters specified in the caption. The yellow region is thus disfavoured at 95\% CL, and we show in dashed the result for a narrow resonance and in solid for $\Gamma/M = 0.3$. 
The fact that the two curves are close shows that the large width effect is not very important at these levels, however we should stress that the region of interest features larger widths than $0.3\ M$, so that care should be taken when extending the projected exclusions.
In magenta we also show the contours with $\Gamma/M = 0.3$ and $\Gamma/M = 1$, showing that the large misalignment region does have larger widths than 30\% of the mass.
In the right side of each plot we show the mass of $h_2$, which is not fixed in the plot but varies with $\ctt$ following
\eq{eq:gamma} 
once we fix $\gamma=0.2$ or $0.4$. The increase of the mass for $\theta \to 0$ is compensated by an increase of the branching ratio into gauge bosons in the same limit, thus explaining why we do not lose too much sensitivity for larger $h_2$ masses.
We also remark that the exclusion crucially depends on $\kappa_g^{h_2}$, thus the yellow regions should be considered as a motivation for further studies rather than actual exclusions.

In this scenario, decays into tops are also relevant:
for instance, in the allowed region of the left panel of \fig{fig:excl-summary}, the branching ratio of $h_2\to t\bar{t}$ lies between $70\%$ and 90\%. However, this is a very challenging search due to large interference between signal and background. Here we have used the framework developed in ref.~\cite{BuarqueFranzosi:2017qlm} to access the power to discover the heavy scalar via top pair production taking these specific parameters as benchmark. 
The analysis is based on the comparison of the measurement of the differential cross section of the $t\bar{t}$ invariant mass distribution in $t\bar{t}$ process at particle level done by the ATLAS collaboration in a resolved~\cite{Aad:2015mbv} and a boosted regime~\cite{Aad:2015hna}. 
Only for large values of $\kappa_g^{h_2}\gtrsim 4\st$ this search becomes competitive with the ZZ search. In the right panel of \fig{fig:excl-summary} the bound from top pair production appears as we chose $\kappa_g^{h_2}=5\st$.
The exclusion is derived by a line-shape analysis on the $m(t\bar{t})$ distribution, which assumes that the data fit exactly the SM prediction for collisions at a centre of mass energy of $\sqrt{s}=13$~TeV and integrated luminosity of 20$\ifb$. We used a $m(t\bar{t})$ resolution of 40 GeV, uncorrelated systematic errors of 15\% on all bins and a theoretical uncertainty of 5\%.  
However, a dedicated experimental analysis searching for this kind of broad resonances, with variable values of the total width and effective gluon couplings, would be necessary to ascertain the reach at the LHC.

\section{Tecni-$\sigma$ interpretation}

As already mentioned, the $\sigma$ discussed in the text can be associated with a light scalar resonance, evidence of which emerge in Lattice studies of theories with an approximate walking behaviour in the Infra-Red (IR). In this section, we summarise the results and also discuss which values for the couplings $k_G'$ and $k_t'$ can be expected in this scenario. We recall that the $\sigma$ under discussion can be identified with the lightest $J^P=0^+$ resonance of the composite dynamics.

The mass of the singlet scalar $0^+$ is  a difficulty quantity to be measured on the lattice~\cite{Brower:2019oor}. Some results are available for a theory based on $\SU(3)$ with fermions in the fundamental, which can play the role of a template for composite Higgs models based on $\SU(4)\times \SU(4)/\SU(4)$ if some of the many flavour are heavy~\cite{Brower:2015owo}. This theory features 12 flavours, 8 of which are heavier while the remaining 4 determine the low energy properties of the theory. Lattice results thus find the presence of a $0^+$ state that remains degenerate with the pNGBs for all the masses probed on the Lattice~\cite{Brower:2015owo,Hasenfratz:2016gut,Aoki:2013zsa}. Similar results have been obtained for 8 light flavours~\cite{Appelquist:2016viq,Appelquist:2018yqe,Aoki:2014oha}, which are believed to be near-IR-conformal (see also ref.~\cite{Aoki:2017fnr}).~\footnote{For recent results in QCD, see ref.~\cite{Briceno:2016mjc}.} 
A light $0^+$ state has also been identified in an $\SU(3)$ model with two Dirac fermions transforming as a sextet~\cite{Fodor:2015vwa} and an $\SU(2)$ model with one Dirac adjoint~\cite{Athenodorou:2014eua}.

The main result contained in these works is that, in theories with enough fermion flavours to be close to an IR conformal fixed point, a light $\sigma$ resonance seems to appear, near-degenerate with the pNGBs. Note that it is very challenging to interpolate this result in the chiral limit, precisely because the value of the light $\sigma$ mass is close to the pion one (which should tend to zero). Nevertheless, the closest results to the chiral limit indicate that the $\sigma$ remains at least lighter than $1/2$ of the mass of the $\rho$. Note that this is not a result applicable to all theories: for instance, for $\SU(2)$ with 2 Dirac fermions in the fundamental, it was found $m_\sigma/f= 19.2(10.8)$~\cite{Arthur:2016dir}, where the large error comes from the difficulty to extract this mass on the Lattice. For an $\SP(4)$ gauge model there are only preliminary results available for a pure glueball state~\cite{Bennett:2017kga}. Progress with dynamical fermions has been reported in refs.~\cite{Lee:2018ztv,Bennett:2019jzz}.

Besides lattice results, information on the $0^+$ mass can be inferred indirectly by its role in unitarising the pNGB scattering amplitude.
From the unitarity of the partial wave amplitudes, the following bound can be extracted:
\beq
m_\sigma < \sqrt{\frac{32 \pi}{N_f}}f \approx \left\{ \begin{array}{c} 7\ f\;\; (N_f=2)\,, \\ 5\ f \;\; (N_f=4)\,, \\ 3.5\ f \;\; (N_f=8)\,. \end{array} \right.
\eeq
Estimates based on Schwinger-Dyson equations, similar to the QCD ones~\cite{Delbourgo:1982tv}, also indicate a low value for $m_\sigma$~\cite{Doff:2019vav}, however the computation has been done for a Technicolor theory with $f = v$ and cannot be extrapolated straightforwardly to our case.
Gravitational dual results also indicate the lightest scalar mass is low compared to the other states~\cite{Elander:2017cle,Elander:2017hyr}.
All these results seem to point to the presence of a light scalar in theories near the conformal window.

We now turn our discussion to the couplings. 
The value of $k_G'$, which can be identified to the coupling of the $\sigma$ to pNGBs, see \eq{eq:sigmapi}, have also been discussed in the literature. For QCD, it has been noticed that the linear sigma model describes amazingly and intriguingly well the data \cite{Belyaev:2013ida}. This corresponds to $k_G'\approx 2$. This coupling has been reproduced on the lattice~\cite{Wang:2017tep}  as well as using dispersion and unitarisation methods~\cite{Pelaez:2010fj}. A similar approximative approach has been used in  CH context in refs.~\cite{Fichet:2016xpw,BuarqueFranzosi:2017prc}. 

\begin{figure}
\centering
\includegraphics[width=0.4\textwidth]{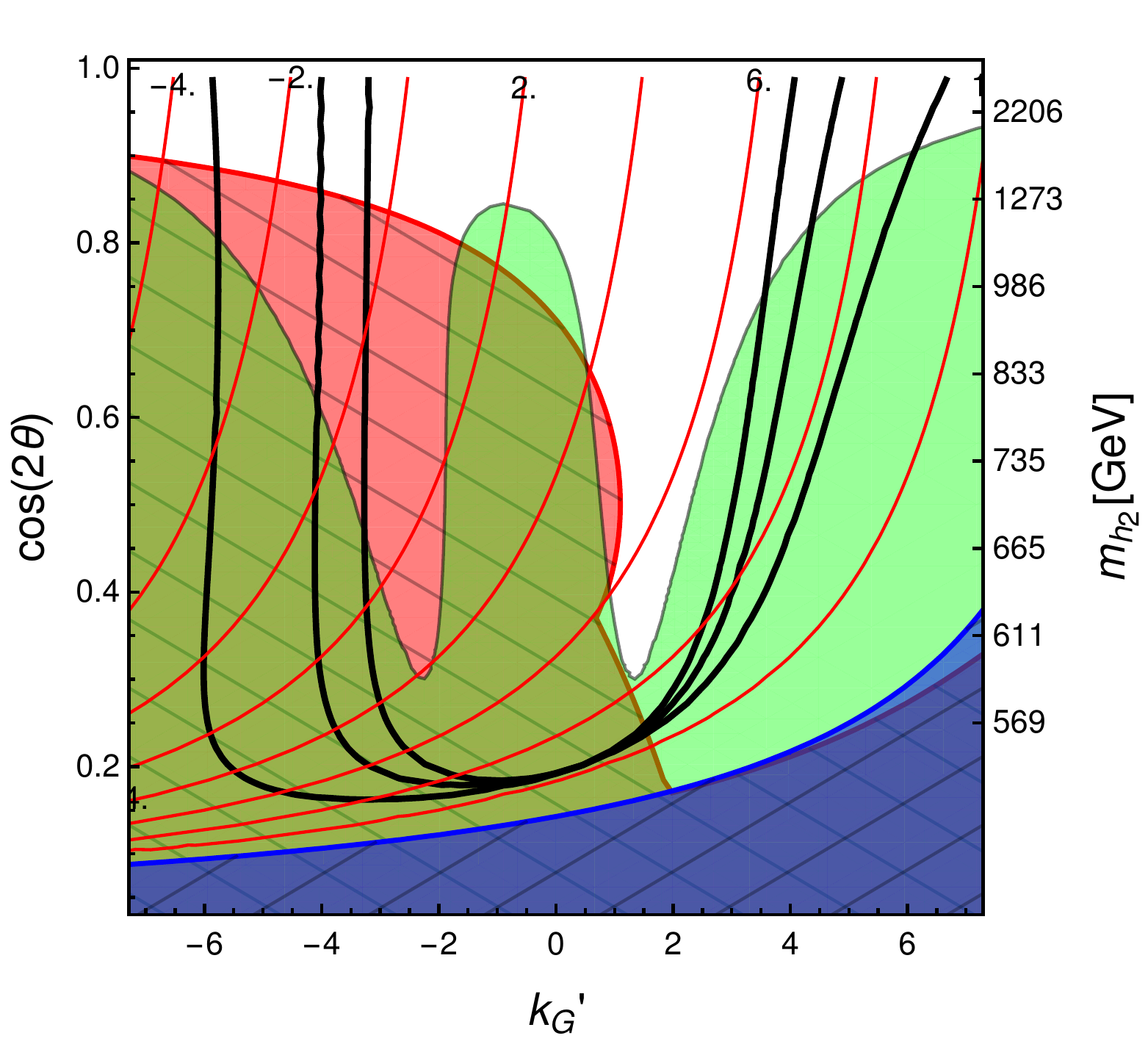}
\caption{Values of $k_t'$ in red contours for $\gamma=0.2$ and $\delta_A=-0.9$. For EWPO we use $\gt=3$ and $r=1.1$.
}
\label{fig:ktp}
\end{figure}

In ref.~\cite{Belyaev:2013ida} the coupling of $\sigma$ to SM fermions was addressed under the assumption of a bilinear giving mass to the fermions \emph{\`a la} Extended Technicolor, which leads to a SM-like coupling $k_t'\approx 1$. This result, however, does not apply to our case where the top mass is generated by partial compositeness. Furthermore, in our case larger $k_t'$ values are necessary to fulfil the considered constraints, as shown in \fig{fig:ktp} for  $\gamma=0.2$ and $\delta_A=- 0.9$. We see that the large misalignment region (low values of $\cos 2\theta$) requires $4\lesssim |k_t'| \lesssim 7$.
Such large couplings would indicate a considerable mixing of the top quark with the top partners. Schematically the origin of the $\sigma \bar{t} t$ coupling can be parameterised as the effect of the mixing of the top with the composite partners, which couple directly to the resonance, yielding (see \eq{eq:kappaT}):
\begin{equation}
k_T' \frac{\sigma}{f}M_T \bar{T} T\;\;  \Rightarrow\;\;  k_T' \frac{M_T \epsilon_{tL} \epsilon_{tR}}{f}\ \sigma \bar{t} t\,,
\end{equation}
with $\epsilon_{tL,R}$ being the mixing of the left and right-handed tops. Therefore, we can estimate
\begin{equation}
k_t' \approx k_T' \frac{M_T \epsilon_{tL} \epsilon_{tR}}{f}\approx \mathcal{O}(4-7)\,,
\end{equation}
which can be achieved with natural coupling values (we recall that $\epsilon_{tL/R} \approx \mathcal{O}(1)$ and $\frac{M_T}{f} \equiv g_T \lesssim 4 \pi$).

In conclusion, we have shown that the $\sigma$ we consider in this work can be matched to the light resonance found in theories with a walking dynamics above the condensation scale.

\section{Conclusions}

We have shown that composite Goldstone Higgs models with compositeness scale as low as $f \approx 400$~GeV are still allowed by low energy date, provided the presence of a sub-TeV Higgs-like resonance. This state is always present in theories with a walking dynamics above the confinement scale, as required by flavour physics. We have shown in a simplified scenario that this result can be achieved rather generally, independently on the details of the coset generating the composite Higgs. Furthermore, the light $\sigma$ state can be associated to the light $0^+$ resonance, evidence of which has been accumulated in recent Lattice studies. The presence of allowed low-scale parameter regions require large values of the couplings of the $\sigma$ to gauge bosons and top quarks, and we have shown that such values are indeed reasonable in dynamical models. We remark that the mass and the couplings of the $\sigma$ are an intrinsic non-tunable property of the underlying dynamics.

This phenomenon, which may occur generically in all composite Higgs models with walking dynamics, predicts the presence of a sub-TeV Higgs-like resonance below $\approx 1$~TeV and with large width, which decays into $ZZ$/$WW$ and $t\bar{t}$. We have shown that the LHC may be able to discover such broad resonance, thus validating or confuting this mechanism, however dedicated large-width searches are needed to provide a definite answer. Finally, Lattice results can provide a further validation of the mechanism by computing the couplings of the resonance to the pNGBs and to the baryonic resonances that couple to the top quark, aka top partners.

Our analysis shows that the common lore that a ``little hierarchy'' between the electroweak scale and the compositeness scale is required should be reconsidered.

\section*{Acknowledgements}

AD and GC acknowledge partial support from the Labex-LIO (Lyon Institute of Origins) 
under grant ANR-10-LABX-66 (Agence Nationale pour la Recherche), and FRAMA 
(FR3127, F\'ed\'eration de Recherche ``Andr\'e Marie Amp\`ere''). DBF acknowledges partial support from the Knut och Alice Wallenberg foundation.

\appendix

\section{Scalar properties} \label{app:scalardetail}

From the Lagrangian in \eq{eq:chiFT}, we can extract the masses, mixing and couplings of the would-be pNGB Higgs $h$ and the singlet $\sigma$.
The mass matrix is given by
\begin{equation}
 - \frac{1}{2} (h \quad \sigma) \left( \begin{array}{cc}
m_h^2 & m_h^2 t_{2\theta}^{-1}(k_G'-k_t')  \\
 m_h^2 t_{2\theta}^{-1}(k_G'-k_t')  & m_\sigma^2
\end{array} \right) \binom{h}{\sigma}\,,
\label{eq:massmat}
\end{equation}
where
\begin{eqnarray}
m_h^2 &=& -8\frac{C_t}{|C_y|^2} m_t^2 \,,  \label{eq:meta}
\end{eqnarray}
while we consider the mass of $\sigma$ as a free parameter as it receives its dominant contribution from the strong dynamics, in the form
$m_\sigma^2 \sim V''_M /f^2 + \mathcal{O} (m_t^2)$. 
As already highlighted, the mass structure above is general and coset-independent.

The masses of the physical eigenstates $h_{1,2}$ are given by
\begin{multline}
m_{h_1,h_2}^2=\frac{1}{2}\left\{ m_\sigma^2+m_h^2 \pm \phantom{\frac{1}{2}} \right. \\
\left. \phantom{\frac{1}{2}} \sqrt{\left(m_\sigma^2-m_h^2\right)^2 + 4(k_G'-k_t')^2 t_{2\theta}^{-2} m_h^4 } \right\}\,.
\end{multline}
In principle, either state can play the role of the $125$~GeV Higgs, however we expect the $\sigma$ to be heavier than the EW scale and its couplings to depart from the SM Higgs ones. Thus, we will conservatively associate the Higgs boson with the lighter state, $m_{h_1} = 125$~GeV, and make sure that in the limit of no mixing, where $(k_G'-k_t')\to 0$, we have $m_{h_1}\to m_{h}$ and $m_{h_2}\to  m_\sigma$. This is justified in the context of a composite sector heavier that the EW scale. The relation between the mass eigenvalues and the mass parameters in the mixing matrix of \eq{eq:massmat} can thus be inverted as follows:
\begin{eqnarray}
m_{h,\sigma}^2 &=& \frac{1}{2[(k_G'-k_t')^2t_{2\theta}^{-2}+1]}\Big\{ m_{h_2}^2+m_{h_1}^2 
   \pm 
   \nonumber \\
   &&\sqrt{(m_{h_2}^2 - m_{h_1}^2)^2-4(k_G'-k_t')^2t_{2\theta}^{-2} m_{h_1}^2m_{h_2}^2 }\Big\} \,.
   \label{eq:mhmsigma}
\end{eqnarray}
We remark that both solutions are real and positive, provided that the argument of the squared root is positive. 
This requires the following relation to hold:
\begin{equation}
\left| \frac{k_G'-k_t'}{t_{2\theta}} \frac{2m_{h_1}m_{h_2}}{m_{h_2}^2-m_{h_1}^2}\right| \leq 1\,.
\end{equation}
This relation explains the definition of $\delta_A$ in \eq{eq:deltaA} and the fact that $|\delta_A|\leq 1$. 
Because $m_{h_1}$ grows monotonically with $m_{h_2}\geq m_{h_1}$, the relation also implies bounds on $m_h$, i.e. the mass term for the pNGB Higgs candidate $h$, reading 
\begin{equation}
m_{h_1}^2 \leq m_{h}^2 \leq 2 \frac{1-\sqrt{1-\delta_A^2}}{\delta_A^2} m_{h_1}^2\,.
\label{eq:mhbounds}
\end{equation}
 In the extreme cases $|\delta_A|=1$, it yields $m_h\leq \sqrt{2}m_{h_1}$.
In \fig{fig:mh} we show values of $m_h$ as a function of $\delta_A$. For $m_{h_2}> 1500\GeV$ the lines fall on top of the others, so the dependence of $m_h$ on $\delta_A$ is quite rigid and $m_h$ quickly saturate to its maximum value in \eq{eq:mhbounds}. 
This result shows that $m_h$ can be larger than the physical value of the Higgs mass, thus contributing on relaxing the tuning associated to the pNGB Higgs mass value.

\begin{figure}
\includegraphics[width=0.4\textwidth]{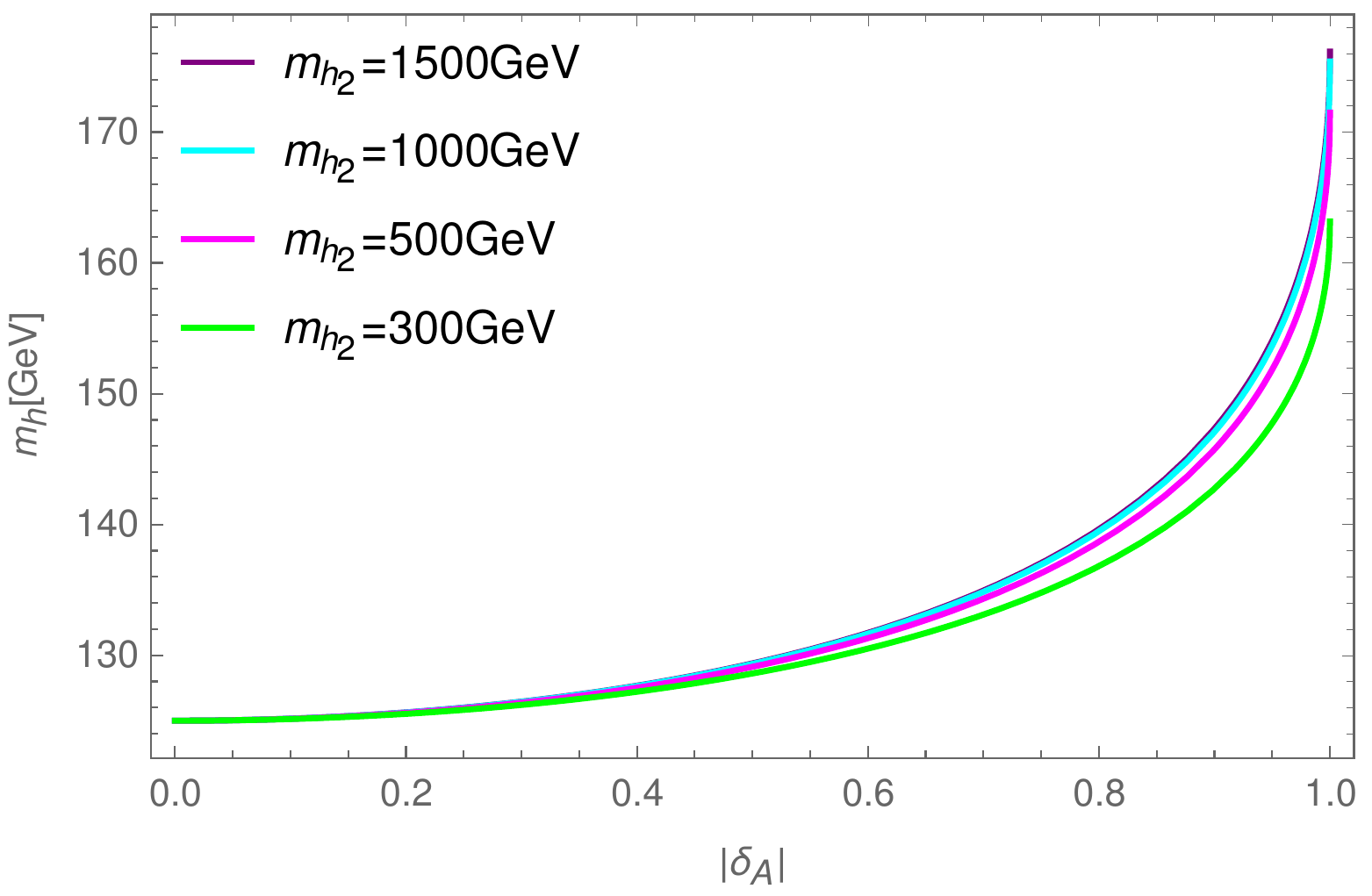}
\caption{$m_h$ as a function of $\delta_A$ for fixed $m_{h_2}$ (and $m_{h_1} = 125$~GeV). }
\label{fig:mh}
\end{figure}

The mass eigenstates are related to $h,\sigma$ by a rotation
\begin{eqnarray}
\binom{h_1}{h_2} =\left( \begin{array}{cc}
c_\alpha & s_\alpha \\
-s_\alpha & c_\alpha 
\end{array} \right) \binom{h}{\sigma} \,,
\end{eqnarray}
with the mixing angle $\alpha$ given by
\begin{eqnarray}
\tan 2\alpha&=&-2\frac{(k_G'-k_t')t_{2\theta}^{-1} m_h^2}{m_\sigma^2-m_h^2}\,.
\end{eqnarray}
The mixing angle has extremal values $t_{2\alpha}=0$ for $\delta_A=0$ and $t_{2\alpha}=\mp 2\frac{m_{h_1}m_{h_2}}{m_{h_2}^2-m_{h_1}^2}\approx \mp 2\frac{m_{h_1}}{m_{h_2}}$ for $\delta_A=\pm 1$, thus it always tends to be small and suppressed by the mass ratio $m_{h_1}/m_{h_2}$. For illustration, in \fig{fig:tan2a} we show some numerical values.

\begin{figure}
\includegraphics[width=0.4\textwidth]{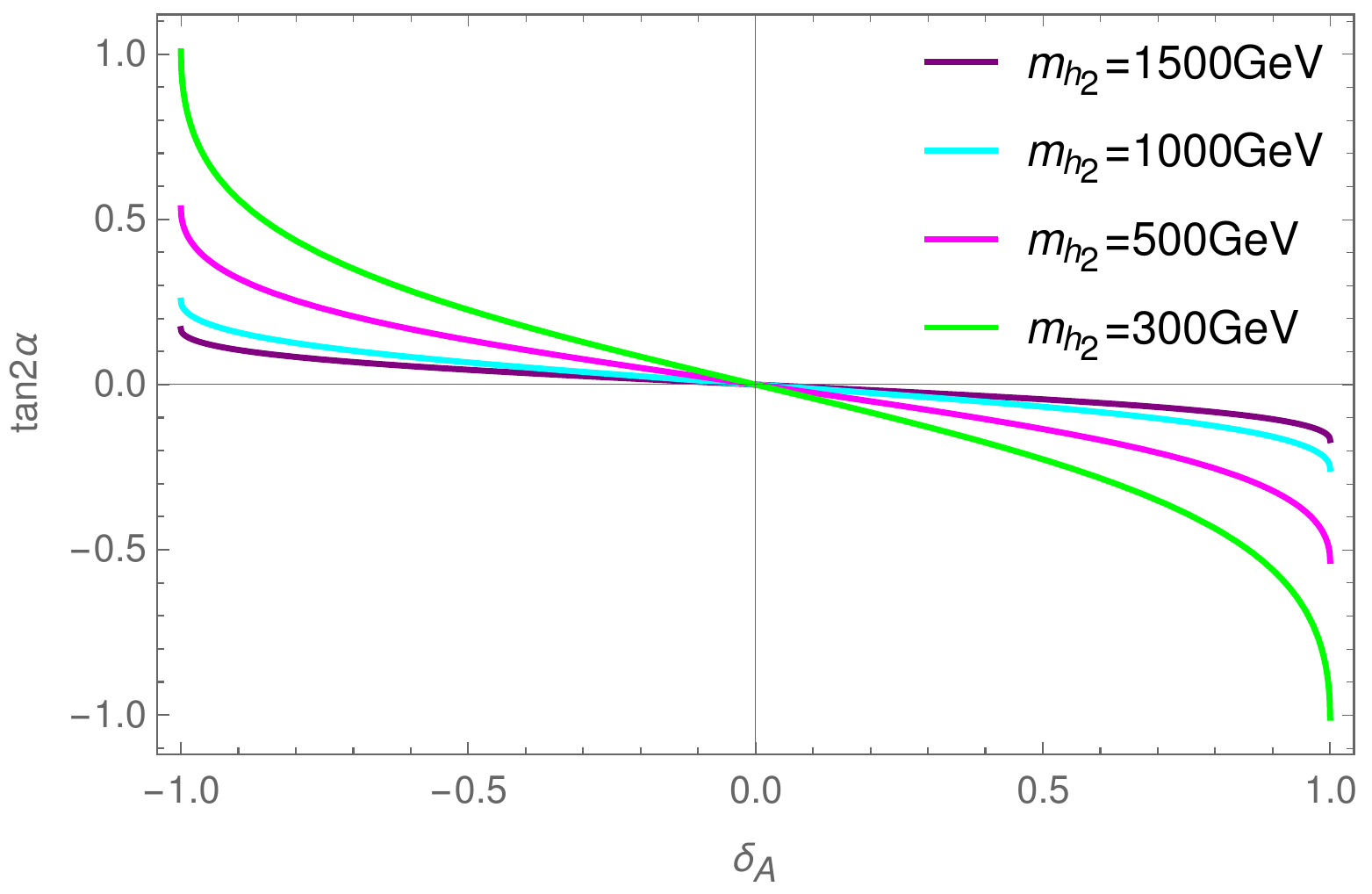}
\caption{$t_{2\alpha}$ as a function of $\delta_A$ for fixed $m_{h_2}$  (and $m_{h_1} = 125$~GeV). }
\label{fig:tan2a}
\end{figure}

\section{Current mass of the underlying fermions and constraints on Higgs BSM branching ratio}
\label{sec:currentmass}

The details provided in the previous section are general as they equally apply to all cosets under study. 
However, the presence of a current mass for the underlying fermions modifies the effective Lagrangian in a different way for different cosets. As it may be relevant to give a mass to all the additional pNGBs,  we will describe its effect here.

The main one is the presence of an additional potential term $- k_m (\sigma) V_{m}$, whose dependence on the misalignment angle depends on the specific coset. 
We always consider here a common mass term for all fermions, aligned with the $\theta=0$ vacuum, knowing that mass differences do not affect the results.
We find that for cosets $\SU(4)/\SP(4)$ and $\SU(4) \times \SU(4)/\SU(4)$ we have
\begin{equation}
V^{\rm A}_m=-4 C_m m_\psi f^3 \ct  + \dots\quad \text{   (case A)}\,,
\end{equation}
while for $\SU(5)/\SO(5)$
\begin{equation}
V^{\rm B}_m =-2 C_m m_\psi f^3 \ct^2  + \dots\,\quad \text{   (case B)}\,,
\end{equation}
where the dots replace terms containing the pNGB fields.
In either form, this term adds 2 extra parameters to our study, $C_m m_\psi$ and $k_m'$, and modifies the vacuum condition and the scalar mass mixing. 
It is convenient to replace the first parameter with the mass term
\begin{equation}
\tilde m_h^2 = -4C_m m_\psi f\,,
\end{equation} 
which provides an additional contribution to the pNGB Higgs mass $h$ equal to  $\tilde m_h^2\ct$ for case A and $\tilde m_h^2\ctt$ for case B.
To avoid cancellations and fine-tuning in the Higgs mass, we will work under the assumption that $\tilde m_h^2 \ll m_{h}^2$, so that the results presented in the main text are only marginally modified. This is consistent with the requirement of generating a small but non-zero mass to some potentially massless pNGBs, as required by phenomenology.

The vacuum condition of \eq{eq:minimum}, i.e. $\delta=\ctt$, is now modified to
\begin{eqnarray}
\tilde m_h^2 &=& 2C_t'f^2 \ct (\ctt-\delta)\quad \text{   (case A)}\,,\\
\tilde m_h^2 &=& C_t'f^2 (\ctt-\delta)\quad \text{   (case B)}  \,.
\end{eqnarray}
The mixing matrix between $h$ and $\sigma$ is now
\begin{equation}
 - \frac{1}{2} (h \quad \sigma) \left( \begin{array}{cc}
m_h^2 & A m_h^2 + B\tilde m_h^2   \\
  A m_h^2 + B\tilde m_h^2  & m_\sigma^2
\end{array} \right) \binom{h}{\sigma}\,,
\label{eq:massmatAB}
\end{equation}
with $A= t_{2\theta}^{-1}(k_G'-k_t')$ as in \eq{eq:massmat} and
\begin{eqnarray}
B&=& \st(k_m'+k_t'-3k_G')+\frac{t_\theta}{2\ct} (k_G'-k_t')\quad \text{(case A)}\,, \\
B&=&\stt(2k_G'-k_m') \quad \text{(case B)}\,.
\end{eqnarray}
$m_h$ is also modified in case A
\begin{equation}
 m_h^2 = -8\frac{C_t}{|C_y|^2} m_t^2 + \frac{\st^2}{\ct}\tilde m_h^2  \quad \text{(case A)}\,,
  \label{eq:mhmhT}
\end{equation}
where the vacuum condition has been used to get rid of $\delta$.

The masses of the physical eigenstates $h_{1,2}$ are given by
\begin{multline}
m_{h_1,h_2}^2=\frac{1}{2}\left\{ m_\sigma^2+m_h^2 \pm \phantom{\frac{1}{2}} \right. \\
\left. \phantom{\frac{1}{2}} \sqrt{\left(m_\sigma^2-m_h^2\right)^2 + 4(Am_h^2+B\tilde{m}_h^2)^2 } \right\}\,.
\end{multline}
We again conservatively associate the Higgs boson with the lighter state, $m_{h_1} = 125$~GeV, and make sure that in the limit of no mixing, where $A,B\to 0$,  $m_h\to m_{h_1}$ and $m_\sigma\to m_{h_2}$. The relation between the mass eigenvalues and the mass parameters in the mixing matrix of \eq{eq:massmat} can be inverted as follows:
\begin{eqnarray}
&&m_{h,\sigma}^2 = \frac{1}{2A^2+1}\Big\{ m_{h_2}^2+m_{h_1}^2-2AB\tilde{m}_h^2 
   \pm 
   \\
  && \sqrt{(m_{h_2}^2+m_{h_1}^2-2AB\tilde{m}_h^2)^2-4(A^2+1)(B^2\tilde{m}_h^2+ m_{h_1}^2m_{h_2}^2) }\Big\} \,.  \nonumber
\end{eqnarray}
This expression reproduces \eq{eq:mhmsigma} for $B=0$. Similarly to the massless fermion case we require the argument of the squared root to be positive and define $\delta_A$ as
\begin{equation}
A= \delta_A \frac{m_{h_2}^2 - m_{h_1}^2}{2 m_{h_2} m_{h_1}} \sqrt{1+\frac{B^2 \tilde{m}_h^4}{m_{h_2}^2 m_{h_1}^2}}
- \frac{m_{h_2}^2 + m_{h_1}^2}{2 m_{h_2} m_{h_1}} \frac{B \tilde{m}_h}{m_{h_2} m_{h_1}}\,,
\end{equation}
and $\delta_A$ is bound to be $-1<\delta_A<1$.

These results show explicitly that the effect of the current mass for the underlying fermions does not have a strong effect on the numerical results we present in the text.

\subsection*{\bf Higgs decay constraint}

The Higgs properties are also affected by the presence of other light pNGBs, whose masses can be below the threshold to contribute to the Higgs decay width.
The global fit of ref.~\cite{Khachatryan:2016vau} provides the following bound on the branching ratio of the Higgs into undetected non-SM states
\begin{equation}
B_{\rm BSM}<0.32\,,
\label{eq:BRbsm}
\end{equation}
at 2 sigmas. 
This imposes a strong constraint on the parameter space due to the Higgs decay.  
We estimate the Higgs total width as
\begin{equation}
\Gamma=(\ct c_\alpha)^2 (\Gamma_b+\Gamma_\tau) + (\kappa_V^{h_1})^2\Gamma_V +\Gamma_g + \Gamma_{\rm BSM}\,,
\end{equation}
where $\Gamma_x$ are the SM partial widths. 
This expression assumes that the bottom and tau get their masses from a bilinear term, as in ref.~\cite{Cacciapaglia:2014uja}. 
For the decay into gluons we use the SM value as a first approximation. 

To compute the BSM partial width $\Gamma_{\rm BSM}$ we consider a particular model base on the coset $\SU(4)/\SP(4)$, where only one extra pseudo-scalar pNGB $\eta$ is present. The decay $h\to \eta\eta$ is driven by the couplings $g_{h\eta^2}$ and $g_{\sigma \eta^2}$ which we computed exactly.
The bound from \eq{eq:BRbsm} is shown as a dashed line in the left panels of 
\fig{fig:constraints} 
 for $m_\eta\approx0$.  
We see that it allows to exclude the large values of $\theta$, thus it is necessary to give mass above threshold to the singlet $m_\eta > m_{h_1}/2$. 
The mass of $\eta$ in this model is given by 
\begin{equation}
m_\eta^2 = \tilde{m}_\eta^2 + \frac{\epsilon_{\bf A}}{4} \left( \frac{m_h^2}{\ct^2} - \frac{\tilde{m}_h^2}{\ct^2} t_\theta\st \right)\,, \label{eq:metasusp}
\end{equation}
where
 $\epsilon_{\bf A} \leq 1$ for the anti-symmetric top spurion, and zero otherwise. 
The actual value of $\epsilon_{\bf A}$ depends on the embedding of the top singlet $t_R$.
Therefore, in the anti-symmetric case, the condition $m_\eta > m_{h_1}/2$ might be fulfilled even for vanishing underlying fermion mass.
Note that Higgs off-shell production  $pp\to h^\ast \to \eta \eta$ might also be sizeable and give interesting final states as $\eta \to Z \gamma$ below the $WW$ threshold~\cite{Galloway:2010bp,Arbey:2015exa}. 
Thus the final state $Z Z \gamma \gamma$ (with off-shell Z if $m_\eta<m_Z$) may be a smoking gun for this model.

\section{Vector meson dominance in EWPOs }
\label{app:EWPO}

To include vector resonances, we construct the Lagrangian using the hidden symmetry technique~\cite{Bando:1987br}.
This method has already been employed in ref.~\cite{Franzosi:2016aoo} for the $\SU(4)/\SP(4)$, and the same results can be used for $\SU(N)/\SP(N)$ and $\SU(N)/\SO(N)$, while the Lagrangian for the QCD-like case $\SU(N)\times \SU(N)/\SU(N)$ can be found in ref.~\cite{Bando:1987br}.

Here we will briefly summarise the key features of this technique in the case of a Higgs coset $G/H$.
We first duplicate the group structure
\begin{equation}
G_0\times G_1/H_0\times H_1\,,
\end{equation}
with two independent sets of NGBs, defined as matrices
\beq
\xi_0 =  e^{i  \Pi_0/f} \qquad \xi_1 =  e^{i \Pi_1/f}
\eeq
that transform nonlinearly as 
\beq
\xi_i\to \xi^\prime_i=g_i \xi_i h(g_i,\pi_i)^\dagger\,,
\eeq
where $g_i$ is an element of $G_i$ and $h$ the corresponding transformation in the subgroup $H_i$. 
We also introduce a new set of NGB from the breaking
\begin{equation}
H_0\times H_1 / H_D\,,
\end{equation}
defined as:
\begin{equation}
	K = \exp\left[ i k^a S^a / f_K \right]\,,
\end{equation}
$S^a$ being generators of $H$, and transforming as
\be
K\to K^\prime=h(g_0,\pi_0)\, K\, h^\dagger(g_1,\pi_1)\,.
\ee

%

To include derivatives and gauge interactions, we define the gauged Maurer-Cartan one-forms as
\begin{equation}
	\omega_{i, \mu}  =  \xi_i^\dag D_{\mu} \xi_i\,,
\end{equation}
where $D_\mu$ are the appropriate covariant derivatives. The core of the hidden symmetry technique is to embed the SM gauge bosons by partly gauging $G_0$, so that
\begin{equation}
D_\mu \xi_0 = ( \partial_\mu -i g {\bf \widetilde{W}}_\mu - i g^\prime {\bf B_\mu} ) \xi_0\,, 
\end{equation}
where ${\bf \widetilde{W}}_\mu$ and ${\bf B_\mu}$ are the gauge bosons of $\SU(2)_L$ and $\UU(1)_Y$ respectively, as embedded in $G_0$. 
They can be written as 
\begin{equation}
{\bf \widetilde{W}}_\mu = W_\mu^a T_L^a\,, \qquad
{\bf B_\mu} = B_\mu T_R^3\,,
\end{equation}
where $T_L^a$ and $T_R^a$ are proportional to generators of $G_0$ ($T_R^a$ for the custodial $\SU(2)_R$).
On the other hand, the vector $\bm{\mathcal{V}}_\mu$ and axial $\bm{\mathcal{A}}_\mu$ resonances are introduced as gauge bosons associated to $G_1$, so that
\begin{equation}
D_\mu \xi_1 = (\partial_\mu -i \widetilde{g} \bm{\mathcal{V}}_\mu - i \widetilde{g} \bm{\mathcal{A}}_\mu )\xi_1\,, 
\end{equation}
where 
\begin{equation}
 \label{eq:vectors0}
\bm{\mathcal{V}}_\mu =  {\cal V}_\mu^a\ \sqrt{2t}\ S_a, \quad \bm{\mathcal{A}}_\mu = {\cal A}_\mu^a\ \sqrt{2t}\ X_a\,,
\end{equation}
where $S^a$ and $X^a$ are the unbroken and broken generators of $G_1/H_1$ respectively (canonically normalised), and $t=\Tr T_LT_L=\Tr T_RT_R$. The normalisation factor $t$ is needed in order to compensate the mismatch in the normalisation of the gauged $T_L$ and $T_R$ generators and the canonically normalised generators of $G/H$.
 The spin-1 fields $\mathcal{V}^j_\mu$ (j=1 to $n_{unbroken}$) and $\mathcal{A}^l_\mu$ (l=1 to $n_{broken}$) are the composite resonances generated by the strong dynamics.

 
To simplify the calculation, we define the broken/unbroken generators with respect to a vacuum that does not contain the Higgs VEV, i.e. $\theta = 0$.
The misalignment angle can thus be re-introduced by rotating the generators associated to the gauging of the EW symmetry. Namely, we define
 \begin{equation}
T_L^i = \Omega \hat{T}_L^i \Omega^\dagger \qquad
T_R^i = \Omega \hat{T}_R^i \Omega^\dagger \,,
\end{equation}
 where $\Omega$ is a rotation of $G$. Now the generators $\hat{T}_{L/R}$ correspond to the custodial symmetry $\SO(4) \sim \SU(2)_L \times \SU(2)_R$ preserved by the $\theta=0$ vacuum (i.e. they are contained in the unbroken generators $S^a$), while the rotation is generated by the Higgs VEV as
 \begin{equation}
\Omega=e^{i\theta X_h/f}\,,
\end{equation}
where $X_h$ (part of the broken generators $X^a$) corresponding to the Higgs component of the doublet that misaligns the vacuum.

 
%
The projections to the broken and unbroken generators are defined respectively by
\begin{eqnarray}
x_{\mu\,i}&=& 2\sum_a \Tr(X_a \omega_{i,\mu})\,X_a\,, \\
v_{\mu\,i}&=&2\sum_a \Tr(S_a \omega_{i,\mu})\,S_a \,,
\end{eqnarray}
so that $v_{\mu\,i}$ transforms inhomogeneously under $G_i$
\be
v_{\mu\,i}\to v_{\mu\,i}^\prime=h(g_i,\pi_i)\, (v_{\mu\,i}+i\partial_\mu)\, h^\dagger(g_i,\pi_i)\,,
\ee
while $x_{\mu\,i}$ transforms homogeneously
\be
x_{\mu\,i}\to x_{\mu\,i}^\prime=h(g_i,\pi_i)\, x_{\mu\,i}\, h^\dagger(g_i,\pi_i)\,
\ee
and can be used to construct invariants for the effective Lagrangian.
%
Note that the $n_{unbroken}$ NGBs contained in $K$ are needed to provide the longitudinal degrees of freedom for the vectors $\mathcal{V}^j_\mu$, while a combination of the NGBs $\pi_i$ from $G_i/H_i$ acts as longitudinal degrees of freedom for the axial $\mathcal{A}^l_\mu$. 
The remaining combination provides the pNGBs of the Higgs coset $G/H$.
%

To lowest order in momentum expansion, 
the effective Lagrangian is given by
\begin{eqnarray}
{\cal L} 
&=&-\frac{1}{4 t}\ {\rm Tr}\ {\bf \widetilde  W}_{\mu\nu} {\bf \widetilde W}^{\mu\nu}
       -\frac{1}{4 t }\ {\rm Tr}\ {\bf B}_{\mu\nu} {\bf B}^{\mu\nu} 
       -\frac{1}{4t}\ {\rm Tr}\ \bm{\mathcal{F}}_{\mu\nu} \bm{\mathcal{F}}^{\mu\nu} \nonumber\\ 
&+&\frac{f_0^2}{N^2}\ {\rm Tr}\ x_{0\mu} x_0^\mu
       + \frac{f_1^2}{N^2}\ {\rm Tr}\ x_{1\mu} x_1^\mu
      +2 r \frac{f_{1}^2}{N^2}\ {\rm Tr}\ x_{0\mu} K x_1^\mu K^\dagger \nonumber \\
&+& \frac{f_K^2}{N^2} \ {\rm Tr}\ {\cal D}^\mu K\ {\cal D}_\mu K^\dagger \,,
\label{eq:DEWSB}
\end{eqnarray}
where the stress-energy tensors are defined, as usual, by
\begin{eqnarray}
{\bf\widetilde{ W}_{\mu\nu}}&=&\partial_\mu {\bf\widetilde{ W}_\nu} - \partial_\nu {\bf\widetilde{ W}_\mu} -\ii g [{\bf\widetilde{ W}_\mu},{\bf\widetilde{ W}_\nu]}\,, \\
{\bf B_{\mu\nu}}&=&\partial_\mu {\bf B_\nu} - \partial_\nu {\bf B_\mu} -\ii g' [{\bf B_\mu},{\bf B_\nu]} \,, \\
\bm{ \mathcal{ F}}_{\mu\nu}&=&\partial_\mu \bm{ \mathcal{ F}}_\nu - \partial_\nu \bm{ \mathcal{ F}}_\mu -\ii \gt [\bm{\mathcal{ F}}_\mu,\bm{\mathcal{ F}}_\nu] \,. 
\end{eqnarray}
Note that the parameter $r$, which plays a crucial role in EWPOs, corresponds to an operator that mixes the SM gauge bosons and the spin-1 resonances, as it multiplies the operator containing the breaking of $H_0\times H_1/H_D$.
The relation between the decay constants $f_0$ and $f_1$ and the EW scale $v=246\GeV$ is fixed to:
\begin{equation}
v^2=\frac{1}{\sqrt{2}G_F}=(f_0^2-f_1^2 r^2)\st^2\,.
\end{equation}

\begin{figure*}
\centering
\includegraphics[width=0.4\textwidth]{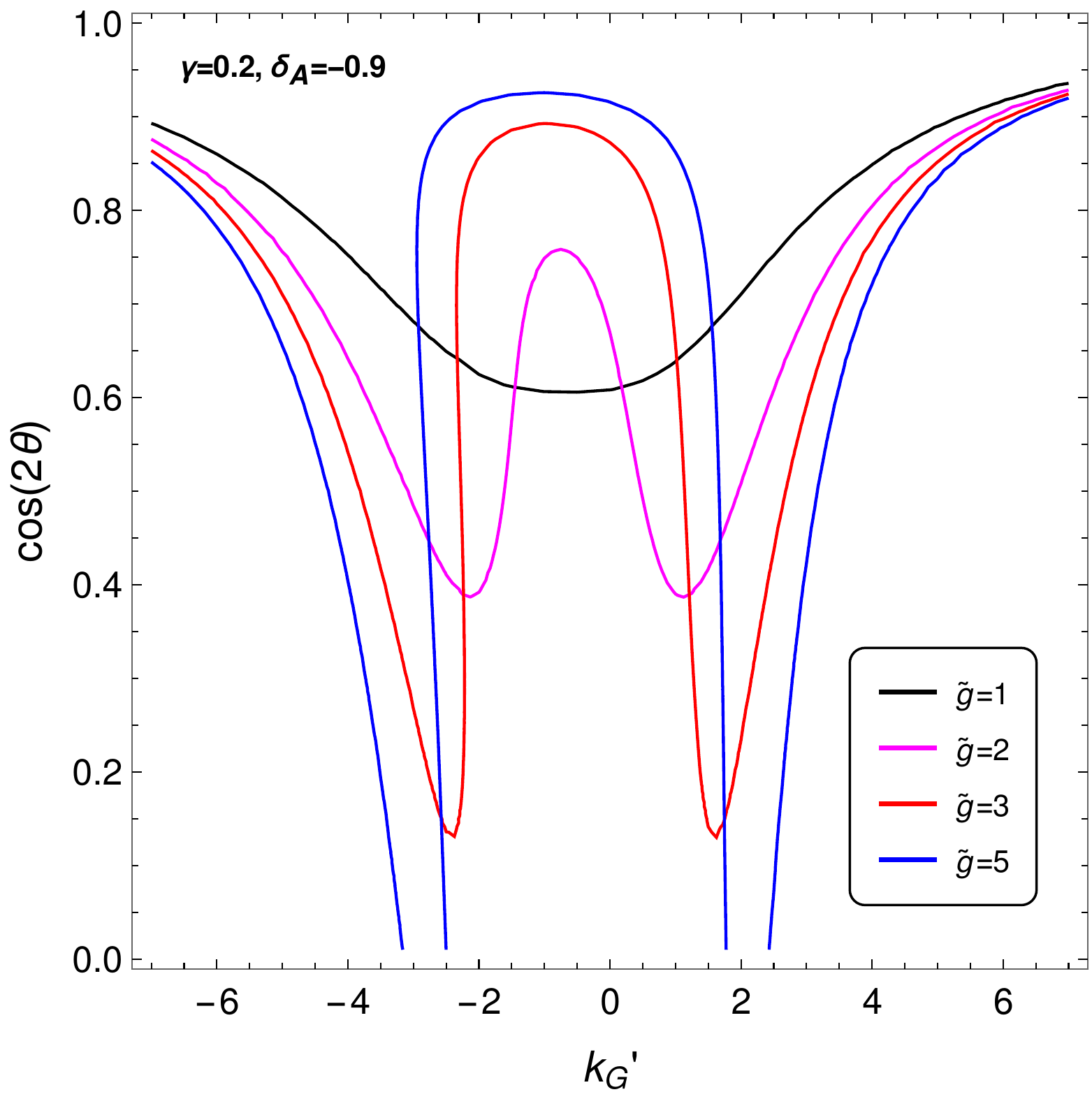}
\includegraphics[width=0.4\textwidth]{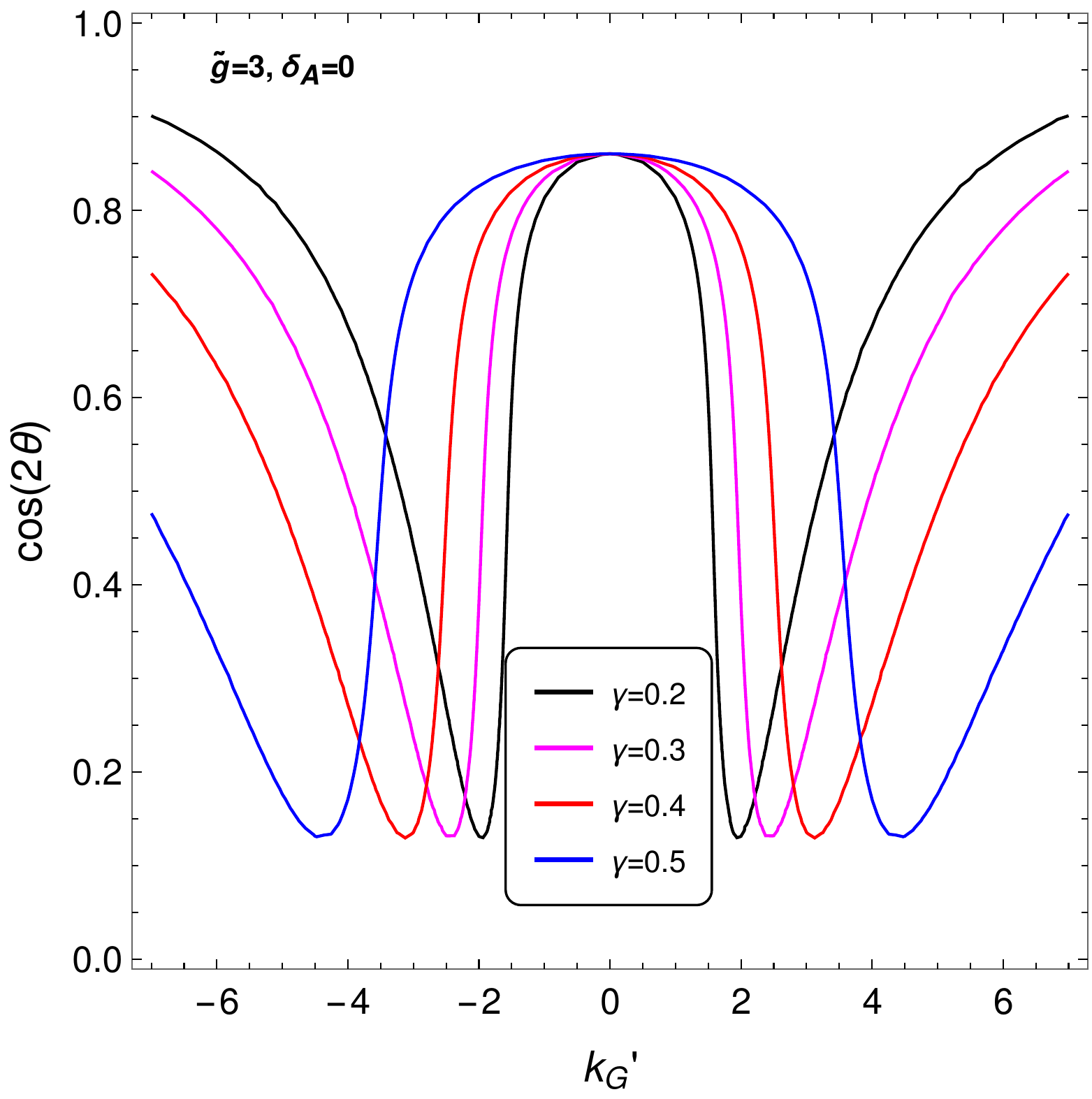}
\caption{EWPO bounds, with $a_\rho=-1$ and $M_\rho=4\pi f$. \emph{Left:} $\gamma=0.2$ and $\delta_A=-0.9$. \emph{Right:} $\gt=3$, $\delta_A=0$. 
}
\label{fig:ewpo}
\end{figure*}

The computation of the oblique parameters follows in a standard way~\cite{Peskin:1990zt,Peskin:1991sw}: 
we first canonically normalise the vector fields and compute the mixing mass matrices that contain the SM fields $B_\mu$ and $W^3_\mu$, $\mathcal{M}_N$,  and $W^\pm_\mu$, $\mathcal{M}_C$. We then proceed by computing the tree-level contributions to the self-energies $\Pi_{WW}$, $\Pi_{33}$, $\Pi_{3B}$ and $\Pi_{BB}$ due to the mixing,
\begin{equation}
\Pi_{WW}(s) = [(s - \mathcal{M}_C)^{-1}]_{11}^{-1}\,, \quad
\Pi_{N}(s) = [(s - \mathcal{M}_N)^{-1}]_{ij}^{-1}\,, 
\end{equation}
with $i,j=1,2$ and $\Pi_{BB}(s)=\Pi_N(s)_{11}$, $\Pi_{3B}(s)=\Pi_N(s)_{12}$ and $\Pi_{33}(s)=\Pi_N(s)_{22}$. 
The $S$ and $T$ parameters are given by
\begin{eqnarray}
S&=&16\pi\left.\frac{g_2\Pi_{3B}'(s)}{g_1 \Pi_{WW}'(s)}\right|_{s=0} \,,\\
T&=&16\pi\left.\frac{\Pi_{33}(s)-\Pi_{WW}(s)}{m_W^2 \Pi_{WW}'(s)}\right|_{s=0} \,.
\end{eqnarray}
By explicit calculation, we obtain the same expression for the 3 cosets under consideration, matching the result given in \eq{eq:rhoewpo}.

This result can be easily understood by tracing back the origin of the correction to the term in the Lagrangian which induces the mixing between the EW gauge bosons, and the massive resonances: the term proportional to $r$.  Because of the traces, the EW gauge bosons can only mix with the vectors and axial-vectors aligned to the same generators in the parallel coset.  The effect of the misalignment is to also involve the spin-1 resonances corresponding to the generators associated to the Higgs doublet that develops a VEV. Thus, as long as the misalignment involves a single doublet, we can always identify the same set of resonances that contribute to $S$ and $T$.
Thus, the only non-universal effects would arise if the misalignment involves more then one direction, besides the Higgs doublet.

In \fig{fig:ewpo} we show how the excluded region from EWPOs varies as a function of some parameters in the model to prove the robustness of the results presented in the main text.
To fix $r$ we choose the unitarity motivated value~\cite{BuarqueFranzosi:2017prc}
\begin{equation}
a_\rho\equiv \frac{\sqrt{2}M_\rho (1-r^2)}{f\gt} =- 1\,,
\end{equation}
 with $M_\rho=4\pi f$. Furthermore, in the \emph{left} panel we fix $\gamma=0.2$ and $\delta_A=-0.9$ and vary $\gt=1,2,3,5$. In the \emph{right} panel we fix $\gt=3$ and $\delta_A=0$ (no mixing) and vary $\gamma$ from 0.2 to 0.5. 
The excluded region in the figure is obtained using a $\chi^2$ method with
$S= 0.04$, $T = 0.08$, $\sigma_S = 0.08$, $\sigma_T = 0.07$ and correlation $\sigma_{TS} = 0.92$~\cite{Haller:2018nnx}. 
The figure shows that the \emph{valleys} always appear for typical values of the parameter space, and they only tend to disappear if $\gt$ is small. However, $\gt$ being the vector coupling, it is expected to be substantially larger than one.
This result again proves the robustness of our general results.


%

\bibliographystyle{JHEP-2-2}

\bibliography{bibliography}

\providecommand{\href}[2]{#2}\begingroup\raggedright\begin{thebibliography}{10}

\bibitem{Bellazzini:2014yua}
B.~Bellazzini, C.~Cs\'aki and J.~Serra, ``{Composite
  Higgses},''\href{http://dx.doi.org/10.1140/epjc/s10052-014-2766-x}{\emph{Eur.
  Phys. J.} {\bf C74} (2014) 2766},
  [\href{https://arxiv.org/abs/1401.2457}{{\tt 1401.2457}}].

\bibitem{Panico:2015jxa}
G.~Panico and A.~Wulzer, ``{The Composite Nambu-Goldstone
  Higgs},''\href{http://dx.doi.org/10.1007/978-3-319-22617-0}{\emph{Lect. Notes
  Phys.} {\bf 913} (2016) pp.1--316},
  [\href{https://arxiv.org/abs/1506.01961}{{\tt 1506.01961}}].

\bibitem{Dugan:1984hq}
M.~J. Dugan, H.~Georgi and D.~B. Kaplan, ``{Anatomy of a Composite Higgs
  Model},''\href{http://dx.doi.org/10.1016/0550-3213(85)90221-4}{\emph{Nucl.
  Phys.} {\bf B254} (1985) 299}.

\bibitem{Kaplan:1983fs}
D.~B. Kaplan and H.~Georgi, ``{SU(2) x U(1) Breaking by Vacuum
  Misalignment},''\href{http://dx.doi.org/10.1016/0370-2693(84)91177-8}{\emph{Phys.
  Lett.} {\bf B136} (1984) 183}.

\bibitem{Weinberg:1975gm}
S.~Weinberg, ``{Implications of Dynamical Symmetry
  Breaking},''\href{http://dx.doi.org/10.1103/PhysRevD.13.974}{\emph{Phys.
  Rev.} {\bf D13} (1976) 974--996}.

\bibitem{Susskind:1978ms}
L.~Susskind, ``{Dynamics of Spontaneous Symmetry Breaking in the Weinberg-Salam
  Theory},''\href{http://dx.doi.org/10.1103/PhysRevD.20.2619}{\emph{Phys. Rev.}
  {\bf D20} (1979) 2619--2625}.

\bibitem{Dimopoulos:1979es}
S.~Dimopoulos and L.~Susskind, ``{Mass Without
  Scalars},''\href{http://dx.doi.org/10.1016/0550-3213(79)90364-X}{\emph{Nucl.
  Phys.} {\bf B155} (1979) 237--252}.

\bibitem{DiVecchia:1980xq}
P.~Di~Vecchia and G.~Veneziano, ``{Minimal Composite Higgs
  Systems},''\href{http://dx.doi.org/10.1016/0370-2693(80)90480-3}{\emph{Phys.
  Lett.} {\bf 95B} (1980) 247--252}.

\bibitem{Dietrich:2005jn}
D.~D. Dietrich, F.~Sannino and K.~Tuominen, ``{Light composite Higgs from
  higher representations versus electroweak precision measurements: Predictions
  for CERN
  LHC},''\href{http://dx.doi.org/10.1103/PhysRevD.72.055001}{\emph{Phys. Rev.}
  {\bf D72} (2005) 055001}, [\href{https://arxiv.org/abs/hep-ph/0505059}{{\tt
  hep-ph/0505059}}].

\bibitem{Belyaev:2013ida}
A.~Belyaev, M.~S. Brown, R.~Foadi and M.~T. Frandsen, ``{The Technicolor Higgs
  in the Light of LHC
  Data},''\href{http://dx.doi.org/10.1103/PhysRevD.90.035012}{\emph{Phys. Rev.}
  {\bf D90} (2014) 035012}, [\href{https://arxiv.org/abs/1309.2097}{{\tt
  1309.2097}}].

\bibitem{Yamawaki:1985zg}
K.~Yamawaki, M.~Bando and K.-i. Matumoto, ``{Scale Invariant Technicolor Model
  and a
  Technidilaton},''\href{http://dx.doi.org/10.1103/PhysRevLett.56.1335}{\emph{Phys.
  Rev. Lett.} {\bf 56} (1986) 1335}.

\bibitem{Bando:1986bg}
M.~Bando, K.-i. Matumoto and K.~Yamawaki,
  ``{Technidilaton},''\href{http://dx.doi.org/10.1016/0370-2693(86)91516-9}{\emph{Phys.
  Lett.} {\bf B178} (1986) 308--312}.

\bibitem{Dzhikiya:1986kk}
G.~V. Dzhikiya, ``{The dilaton as the analog of the Higgs boson in composite
  models},''{\emph{Sov. J. Nucl. Phys.} {\bf 45} (1987) 1083--1087}.

\bibitem{Peskin:1990zt}
M.~E. Peskin and T.~Takeuchi, ``{A New constraint on a strongly interacting
  Higgs
  sector},''\href{http://dx.doi.org/10.1103/PhysRevLett.65.964}{\emph{Phys.
  Rev. Lett.} {\bf 65} (1990) 964--967}.

\bibitem{Peskin:1991sw}
M.~E. Peskin and T.~Takeuchi, ``{Estimation of oblique electroweak
  corrections},''\href{http://dx.doi.org/10.1103/PhysRevD.46.381}{\emph{Phys.
  Rev.} {\bf D46} (1992) 381--409}.

\bibitem{Barbieri:2004qk}
R.~Barbieri, A.~Pomarol, R.~Rattazzi and A.~Strumia, ``{Electroweak symmetry
  breaking after LEP-1 and
  LEP-2},''\href{http://dx.doi.org/10.1016/j.nuclphysb.2004.10.014}{\emph{Nucl.
  Phys.} {\bf B703} (2004) 127--146},
  [\href{https://arxiv.org/abs/hep-ph/0405040}{{\tt hep-ph/0405040}}].

\bibitem{Barbieri:2012tu}
R.~Barbieri, D.~Buttazzo, F.~Sala, D.~M. Straub and A.~Tesi, ``{A 125 GeV
  composite Higgs boson versus flavour and electroweak precision
  tests},''\href{http://dx.doi.org/10.1007/JHEP05(2013)069}{\emph{JHEP} {\bf
  05} (2013) 069}, [\href{https://arxiv.org/abs/1211.5085}{{\tt 1211.5085}}].

\bibitem{Grojean:2013qca}
C.~Grojean, O.~Matsedonskyi and G.~Panico, ``{Light top partners and precision
  physics},''\href{http://dx.doi.org/10.1007/JHEP10(2013)160}{\emph{JHEP} {\bf
  10} (2013) 160}, [\href{https://arxiv.org/abs/1306.4655}{{\tt 1306.4655}}].

\bibitem{Arbey:2015exa}
A.~Arbey, G.~Cacciapaglia, H.~Cai, A.~Deandrea, S.~Le~Corre and F.~Sannino,
  ``{Fundamental Composite Electroweak Dynamics: Status at the
  LHC},''\href{http://dx.doi.org/10.1103/PhysRevD.95.015028}{\emph{Phys. Rev.}
  {\bf D95} (2017) 015028}, [\href{https://arxiv.org/abs/1502.04718}{{\tt
  1502.04718}}].

\bibitem{Matsedonskyi:2012ym}
O.~Matsedonskyi, G.~Panico and A.~Wulzer, ``{Light Top Partners for a Light
  Composite
  Higgs},''\href{http://dx.doi.org/10.1007/JHEP01(2013)164}{\emph{JHEP} {\bf
  01} (2013) 164}, [\href{https://arxiv.org/abs/1204.6333}{{\tt 1204.6333}}].

\bibitem{Galloway:2010bp}
J.~Galloway, J.~A. Evans, M.~A. Luty and R.~A. Tacchi, ``{Minimal Conformal
  Technicolor and Precision Electroweak
  Tests},''\href{http://dx.doi.org/10.1007/JHEP10(2010)086}{\emph{JHEP} {\bf
  10} (2010) 086}, [\href{https://arxiv.org/abs/1001.1361}{{\tt 1001.1361}}].

\bibitem{Cacciapaglia:2014uja}
G.~Cacciapaglia and F.~Sannino, ``{Fundamental Composite (Goldstone) Higgs
  Dynamics},''\href{http://dx.doi.org/10.1007/JHEP04(2014)111}{\emph{JHEP} {\bf
  1404} (2014) 111}, [\href{https://arxiv.org/abs/1402.0233}{{\tt 1402.0233}}].

\bibitem{Csaki:2017cep}
C.~Cs\'aki, T.~Ma and J.~Shu, ``{Maximally Symmetric Composite Higgs
  Models},''\href{http://dx.doi.org/10.1103/PhysRevLett.119.131803}{\emph{Phys.
  Rev. Lett.} {\bf 119} (2017) 131803},
  [\href{https://arxiv.org/abs/1702.00405}{{\tt 1702.00405}}].

\bibitem{Csaki:2017jby}
C.~Cs\'aki, T.~Ma and J.~Shu, ``{Trigonometric Parity for the Composite
  Higgs},'' \href{https://arxiv.org/abs/1709.08636}{{\tt 1709.08636}}.

\bibitem{Dimopoulos:134187}
S.~K. Dimopoulos and J.~R. Ellis, ``{Challenges for extended technicolour
  theories},''{\emph{Nucl. Phys. B} {\bf 182} (Sep, 1980) 505--528. 31 p}.

\bibitem{Hasenfratz:2016gut}
A.~Hasenfratz, C.~Rebbi and O.~Witzel, ``{Large scale separation and resonances
  within LHC range from a prototype BSM
  model},''\href{http://dx.doi.org/10.1016/j.physletb.2017.07.058}{\emph{Phys.
  Lett.} {\bf B773} (2017) 86--90},
  [\href{https://arxiv.org/abs/1609.01401}{{\tt 1609.01401}}].

\bibitem{Athenodorou:2016ndx}
A.~Athenodorou, E.~Bennett, G.~Bergner, D.~Elander, C.~J.~D. Lin, B.~Lucini
  et~al., ``{Large mass hierarchies from strongly-coupled
  dynamics},''\href{http://dx.doi.org/10.1007/JHEP06(2016)114}{\emph{JHEP} {\bf
  06} (2016) 114}, [\href{https://arxiv.org/abs/1605.04258}{{\tt 1605.04258}}].

\bibitem{Aoki:2017fnr}
Y.~Aoki et~al., ``{Flavor-singlet spectrum in multi-flavor
  QCD},''\href{http://dx.doi.org/10.1051/epjconf/201817508023}{\emph{EPJ Web
  Conf.} {\bf 175} (2018) 08023}, [\href{https://arxiv.org/abs/1710.06549}{{\tt
  1710.06549}}].

\bibitem{Appelquist:2018yqe}
{\scshape Lattice Strong Dynamics} collaboration, T.~Appelquist et~al.,
  ``{Nonperturbative investigations of SU(3) gauge theory with eight dynamical
  flavors},''\href{http://dx.doi.org/10.1103/PhysRevD.99.014509}{\emph{Phys.
  Rev.} {\bf D99} (2019) 014509}, [\href{https://arxiv.org/abs/1807.08411}{{\tt
  1807.08411}}].

\bibitem{Elander:2017cle}
D.~Elander and M.~Piai, ``{Calculable mass hierarchies and a light dilaton from
  gravity
  duals},''\href{http://dx.doi.org/10.1016/j.physletb.2017.06.035}{\emph{Phys.
  Lett.} {\bf B772} (2017) 110--114},
  [\href{https://arxiv.org/abs/1703.09205}{{\tt 1703.09205}}].

\bibitem{Elander:2017hyr}
D.~Elander and M.~Piai, ``{Glueballs on the Baryonic Branch of
  Klebanov-Strassler: dimensional deconstruction and a light scalar
  particle},''\href{http://dx.doi.org/10.1007/JHEP06(2017)003}{\emph{JHEP} {\bf
  06} (2017) 003}, [\href{https://arxiv.org/abs/1703.10158}{{\tt 1703.10158}}].

\bibitem{BitaghsirFadafan:2018efw}
K.~Bitaghsir~Fadafan, W.~Clemens and N.~Evans, ``{Holographic Gauged NJL Model:
  the Conformal Window and Ideal
  Walking},''\href{http://dx.doi.org/10.1103/PhysRevD.98.066015}{\emph{Phys.
  Rev.} {\bf D98} (2018) 066015}, [\href{https://arxiv.org/abs/1807.04548}{{\tt
  1807.04548}}].

\bibitem{Holdom:1981rm}
B.~Holdom, ``{Raising the Sideways
  Scale},''\href{http://dx.doi.org/10.1103/PhysRevD.24.1441}{\emph{Phys. Rev.}
  {\bf D24} (1981) 1441}.

\bibitem{Matsedonskyi:2014iha}
O.~Matsedonskyi, ``{On Flavour and Naturalness of Composite Higgs
  Models},''\href{http://dx.doi.org/10.1007/JHEP02(2015)154}{\emph{JHEP} {\bf
  02} (2015) 154}, [\href{https://arxiv.org/abs/1411.4638}{{\tt 1411.4638}}].

\bibitem{Cacciapaglia:2015dsa}
G.~Cacciapaglia, H.~Cai, T.~Flacke, S.~J. Lee, A.~Parolini and H.~Ser\^odio,
  ``{Anarchic Yukawas and top partial compositeness: the flavour of a
  successful
  marriage},''\href{http://dx.doi.org/10.1007/JHEP06(2015)085}{\emph{JHEP} {\bf
  06} (2015) 085}, [\href{https://arxiv.org/abs/1501.03818}{{\tt 1501.03818}}].

\bibitem{Kaplan:1991dc}
D.~B. Kaplan, ``{Flavor at SSC energies: A New mechanism for dynamically
  generated fermion
  masses},''\href{http://dx.doi.org/10.1016/S0550-3213(05)80021-5}{\emph{Nucl.
  Phys.} {\bf B365} (1991) 259--278}.

\bibitem{Golterman:2017vdj}
M.~Golterman and Y.~Shamir, ``{Effective potential in ultraviolet completions
  for composite Higgs
  models},''\href{http://dx.doi.org/10.1103/PhysRevD.97.095005}{\emph{Phys.
  Rev.} {\bf D97} (2018) 095005}, [\href{https://arxiv.org/abs/1707.06033}{{\tt
  1707.06033}}].

\bibitem{Alanne:2018wtp}
T.~Alanne, N.~Bizot, G.~Cacciapaglia and F.~Sannino, ``{Classification of NLO
  operators for composite Higgs
  models},''\href{http://dx.doi.org/10.1103/PhysRevD.97.075028}{\emph{Phys.
  Rev.} {\bf D97} (2018) 075028}, [\href{https://arxiv.org/abs/1801.05444}{{\tt
  1801.05444}}].

\bibitem{Agugliaro:2018vsu}
A.~Agugliaro, G.~Cacciapaglia, A.~Deandrea and S.~De~Curtis, ``{Vacuum
  misalignment and pattern of scalar masses in the SU(5)/SO(5) composite Higgs
  model},'' \href{https://arxiv.org/abs/1808.10175}{{\tt 1808.10175}}.

\bibitem{Liu:2018vel}
D.~Liu, I.~Low and Z.~Yin, ``{Universal Imprints of a Pseudo-Nambu-Goldstone
  Higgs
  Boson},''\href{http://dx.doi.org/10.1103/PhysRevLett.121.261802}{\emph{Phys.
  Rev. Lett.} {\bf 121} (2018) 261802},
  [\href{https://arxiv.org/abs/1805.00489}{{\tt 1805.00489}}].

\bibitem{Sirunyan:2018hoz}
{\scshape CMS} collaboration, A.~M. Sirunyan et~al., ``{Observation of
  $\mathrm{t\overline{t}}$H
  production},''\href{http://dx.doi.org/10.1103/PhysRevLett.120.231801,
  10.1130/PhysRevLett.120.231801}{\emph{Phys. Rev. Lett.} {\bf 120} (2018)
  231801}, [\href{https://arxiv.org/abs/1804.02610}{{\tt 1804.02610}}].

\bibitem{Aaboud:2018urx}
{\scshape ATLAS} collaboration, M.~Aaboud et~al., ``{Observation of Higgs boson
  production in association with a top quark pair at the LHC with the ATLAS
  detector},''\href{http://dx.doi.org/10.1016/j.physletb.2018.07.035}{\emph{Phys.
  Lett.} {\bf B784} (2018) 173--191},
  [\href{https://arxiv.org/abs/1806.00425}{{\tt 1806.00425}}].

\bibitem{Hansen:2016fri}
M.~Hansen, K.~Lang{\ae}ble and F.~Sannino, ``{Extending Chiral Perturbation
  Theory with an Isosinglet
  Scalar},''\href{http://dx.doi.org/10.1103/PhysRevD.95.036005}{\emph{Phys.
  Rev.} {\bf D95} (2017) 036005}, [\href{https://arxiv.org/abs/1610.02904}{{\tt
  1610.02904}}].

\bibitem{Golterman:2016lsd}
M.~Golterman and Y.~Shamir, ``{Low-energy effective action for pions and a
  dilatonic
  meson},''\href{http://dx.doi.org/10.1103/PhysRevD.94.054502}{\emph{Phys.
  Rev.} {\bf D94} (2016) 054502}, [\href{https://arxiv.org/abs/1603.04575}{{\tt
  1603.04575}}].

\bibitem{Appelquist:2017wcg}
T.~Appelquist, J.~Ingoldby and M.~Piai, ``{Dilaton EFT Framework For Lattice
  Data},''\href{http://dx.doi.org/10.1007/JHEP07(2017)035}{\emph{JHEP} {\bf 07}
  (2017) 035}, [\href{https://arxiv.org/abs/1702.04410}{{\tt 1702.04410}}].

\bibitem{Ayyar:2019exp}
V.~Ayyar, M.~F. Golterman, D.~C. Hackett, W.~Jay, E.~T. Neil, Y.~Shamir et~al.,
  ``{Radiative Contribution to the Composite-Higgs Potential in a
  Two-Representation Lattice
  Model},''\href{http://dx.doi.org/10.1103/PhysRevD.99.094504}{\emph{Phys.
  Rev.} {\bf D99} (2019) 094504}, [\href{https://arxiv.org/abs/1903.02535}{{\tt
  1903.02535}}].

\bibitem{BuarqueFranzosi:2017prc}
D.~Buarque~Franzosi and P.~Ferrarese, ``{Implications of Vector Boson
  Scattering Unitarity in Composite Higgs
  Models},''\href{http://dx.doi.org/10.1103/PhysRevD.96.055037}{\emph{Phys.
  Rev.} {\bf D96} (2017) 055037}, [\href{https://arxiv.org/abs/1705.02787}{{\tt
  1705.02787}}].

\bibitem{Khachatryan:2016vau}
{\scshape ATLAS, CMS} collaboration, G.~Aad et~al., ``{Measurements of the
  Higgs boson production and decay rates and constraints on its couplings from
  a combined ATLAS and CMS analysis of the LHC pp collision data at $
  \sqrt{s}=7 $ and 8
  TeV},''\href{http://dx.doi.org/10.1007/JHEP08(2016)045}{\emph{JHEP} {\bf 08}
  (2016) 045}, [\href{https://arxiv.org/abs/1606.02266}{{\tt 1606.02266}}].

\bibitem{Haller:2018nnx}
J.~Haller, A.~Hoecker, R.~Kogler, K.~Mönig, T.~Peiffer and J.~Stelzer,
  ``{Update of the global electroweak fit and constraints on two-Higgs-doublet
  models},''\href{http://dx.doi.org/10.1140/epjc/s10052-018-6131-3}{\emph{Eur.
  Phys. J.} {\bf C78} (2018) 675},
  [\href{https://arxiv.org/abs/1803.01853}{{\tt 1803.01853}}].

\bibitem{Harada:1995dc}
M.~Harada, F.~Sannino and J.~Schechter, ``{Simple description of pi pi
  scattering to
  1-GeV},''\href{http://dx.doi.org/10.1103/PhysRevD.54.1991}{\emph{Phys. Rev.}
  {\bf D54} (1996) 1991--2004},
  [\href{https://arxiv.org/abs/hep-ph/9511335}{{\tt hep-ph/9511335}}].

\bibitem{Sannino:2010ca}
F.~Sannino, ``{Mass Deformed Exact S-parameter in Conformal
  Theories},''\href{http://dx.doi.org/10.1103/PhysRevD.82.081701}{\emph{Phys.
  Rev.} {\bf D82} (2010) 081701}, [\href{https://arxiv.org/abs/1006.0207}{{\tt
  1006.0207}}].

\bibitem{Casalbuoni:1995yb}
R.~Casalbuoni, A.~Deandrea, S.~De~Curtis, D.~Dominici, F.~Feruglio, R.~Gatto
  et~al., ``{Symmetries for vector and axial vector
  mesons},''\href{http://dx.doi.org/10.1016/0370-2693(95)00291-R}{\emph{Phys.
  Lett.} {\bf B349} (1995) 533--540},
  [\href{https://arxiv.org/abs/hep-ph/9502247}{{\tt hep-ph/9502247}}].

\bibitem{Contino:2015mha}
R.~Contino and M.~Salvarezza, ``{One-loop effects from spin-1 resonances in
  Composite Higgs
  models},''\href{http://dx.doi.org/10.1007/JHEP07(2015)065}{\emph{JHEP} {\bf
  07} (2015) 065}, [\href{https://arxiv.org/abs/1504.02750}{{\tt 1504.02750}}].

\bibitem{Franzosi:2016aoo}
D.~Buarque~Franzosi, G.~Cacciapaglia, H.~Cai, A.~Deandrea and M.~Frandsen,
  ``{Vector and Axial-vector resonances in composite models of the Higgs
  boson},''\href{http://dx.doi.org/10.1007/JHEP11(2016)076}{\emph{JHEP} {\bf
  11} (2016) 076}, [\href{https://arxiv.org/abs/1605.01363}{{\tt 1605.01363}}].

\bibitem{Ma:2015gra}
T.~Ma and G.~Cacciapaglia, ``{Fundamental Composite 2HDM: SU(N) with 4
  flavours},''\href{http://dx.doi.org/10.1007/JHEP03(2016)211}{\emph{JHEP} {\bf
  03} (2016) 211}, [\href{https://arxiv.org/abs/1508.07014}{{\tt 1508.07014}}].

\bibitem{Sirunyan:2018qlb}
{\scshape CMS} collaboration, A.~M. Sirunyan et~al., ``{Search for a new scalar
  resonance decaying to a pair of Z bosons in proton-proton collisions at
  $\sqrt{s}=13 $
  TeV},''\href{http://dx.doi.org/10.1007/JHEP06(2018)127}{\emph{JHEP} {\bf 06}
  (2018) 127}, [\href{https://arxiv.org/abs/1804.01939}{{\tt 1804.01939}}].

\bibitem{Anastasiou:2016hlm}
C.~Anastasiou, C.~Duhr, F.~Dulat, E.~Furlan, T.~Gehrmann, F.~Herzog et~al.,
  ``{CP-even scalar boson production via gluon fusion at the
  LHC},''\href{http://dx.doi.org/10.1007/JHEP09(2016)037}{\emph{JHEP} {\bf 09}
  (2016) 037}, [\href{https://arxiv.org/abs/1605.05761}{{\tt 1605.05761}}].

\bibitem{Bolzoni:2010xr}
P.~Bolzoni, F.~Maltoni, S.-O. Moch and M.~Zaro, ``{Higgs production via
  vector-boson fusion at NNLO in
  QCD},''\href{http://dx.doi.org/10.1103/PhysRevLett.105.011801}{\emph{Phys.
  Rev. Lett.} {\bf 105} (2010) 011801},
  [\href{https://arxiv.org/abs/1003.4451}{{\tt 1003.4451}}].

\bibitem{Bolzoni:2011cu}
P.~Bolzoni, F.~Maltoni, S.-O. Moch and M.~Zaro, ``{Vector boson fusion at NNLO
  in QCD: SM Higgs and
  beyond},''\href{http://dx.doi.org/10.1103/PhysRevD.85.035002}{\emph{Phys.
  Rev.} {\bf D85} (2012) 035002}, [\href{https://arxiv.org/abs/1109.3717}{{\tt
  1109.3717}}].

\bibitem{BuarqueFranzosi:2017qlm}
D.~Buarque~Franzosi, F.~Fabbri and S.~Schumann, ``{Constraining scalar
  resonances with top-quark pair production at the
  LHC},''\href{http://dx.doi.org/10.1007/JHEP03(2018)022}{\emph{JHEP} {\bf 03}
  (2018) 022}, [\href{https://arxiv.org/abs/1711.00102}{{\tt 1711.00102}}].

\bibitem{Aad:2015mbv}
{\scshape ATLAS} collaboration, G.~Aad et~al., ``{Measurements of top-quark
  pair differential cross-sections in the lepton+jets channel in $pp$
  collisions at $\sqrt{s}=8$ TeV using the ATLAS
  detector},''\href{http://dx.doi.org/10.1140/epjc/s10052-016-4366-4}{\emph{Eur.
  Phys. J.} {\bf C76} (2016) 538},
  [\href{https://arxiv.org/abs/1511.04716}{{\tt 1511.04716}}].

\bibitem{Aad:2015hna}
{\scshape ATLAS} collaboration, G.~Aad et~al., ``{Measurement of the
  differential cross-section of highly boosted top quarks as a function of
  their transverse momentum in $\sqrt{s}$ = 8 TeV proton-proton collisions
  using the ATLAS
  detector},''\href{http://dx.doi.org/10.1103/PhysRevD.93.032009}{\emph{Phys.
  Rev.} {\bf D93} (2016) 032009}, [\href{https://arxiv.org/abs/1510.03818}{{\tt
  1510.03818}}].

\bibitem{Brower:2019oor}
{\scshape USQCD} collaboration, R.~C. Brower, A.~Hasenfratz, E.~T. Neil,
  S.~Catterall, G.~Fleming, J.~Giedt et~al., ``{Lattice Gauge Theory for
  Physics Beyond the Standard Model},''
  \href{https://arxiv.org/abs/1904.09964}{{\tt 1904.09964}}.

\bibitem{Brower:2015owo}
R.~C. Brower, A.~Hasenfratz, C.~Rebbi, E.~Weinberg and O.~Witzel, ``{Composite
  Higgs model at a conformal fixed
  point},''\href{http://dx.doi.org/10.1103/PhysRevD.93.075028}{\emph{Phys.
  Rev.} {\bf D93} (2016) 075028}, [\href{https://arxiv.org/abs/1512.02576}{{\tt
  1512.02576}}].

\bibitem{Aoki:2013zsa}
{\scshape LatKMI} collaboration, Y.~Aoki, T.~Aoyama, M.~Kurachi, T.~Maskawa,
  K.-i. Nagai, H.~Ohki et~al., ``{Light composite scalar in twelve-flavor QCD
  on the
  lattice},''\href{http://dx.doi.org/10.1103/PhysRevLett.111.162001}{\emph{Phys.
  Rev. Lett.} {\bf 111} (2013) 162001},
  [\href{https://arxiv.org/abs/1305.6006}{{\tt 1305.6006}}].

\bibitem{Appelquist:2016viq}
T.~Appelquist et~al., ``{Strongly interacting dynamics and the search for new
  physics at the
  LHC},''\href{http://dx.doi.org/10.1103/PhysRevD.93.114514}{\emph{Phys. Rev.}
  {\bf D93} (2016) 114514}, [\href{https://arxiv.org/abs/1601.04027}{{\tt
  1601.04027}}].

\bibitem{Aoki:2014oha}
{\scshape LatKMI} collaboration, Y.~Aoki et~al., ``{Light composite scalar in
  eight-flavor QCD on the
  lattice},''\href{http://dx.doi.org/10.1103/PhysRevD.89.111502}{\emph{Phys.
  Rev.} {\bf D89} (2014) 111502}, [\href{https://arxiv.org/abs/1403.5000}{{\tt
  1403.5000}}].

\bibitem{Briceno:2016mjc}
R.~A. Briceno, J.~J. Dudek, R.~G. Edwards and D.~J. Wilson, ``{Isoscalar
  $\pi\pi$ scattering and the $\sigma$ meson resonance from
  QCD},''\href{http://dx.doi.org/10.1103/PhysRevLett.118.022002}{\emph{Phys.
  Rev. Lett.} {\bf 118} (2017) 022002},
  [\href{https://arxiv.org/abs/1607.05900}{{\tt 1607.05900}}].

\bibitem{Fodor:2015vwa}
Z.~Fodor, K.~Holland, J.~Kuti, S.~Mondal, D.~Nogradi and C.~H. Wong, ``{Toward
  the minimal realization of a light composite
  Higgs},''\href{http://dx.doi.org/10.22323/1.214.0244}{\emph{PoS} {\bf
  LATTICE2014} (2015) 244}, [\href{https://arxiv.org/abs/1502.00028}{{\tt
  1502.00028}}].

\bibitem{Athenodorou:2014eua}
A.~Athenodorou, E.~Bennett, G.~Bergner and B.~Lucini, ``{Infrared regime of
  SU(2) with one adjoint Dirac
  flavor},''\href{http://dx.doi.org/10.1103/PhysRevD.91.114508}{\emph{Phys.
  Rev.} {\bf D91} (2015) 114508}, [\href{https://arxiv.org/abs/1412.5994}{{\tt
  1412.5994}}].

\bibitem{Arthur:2016dir}
R.~Arthur, V.~Drach, M.~Hansen, A.~Hietanen, C.~Pica and F.~Sannino, ``{SU(2)
  Gauge Theory with Two Fundamental Flavours: a Minimal Template for Model
  Building},'' \href{https://arxiv.org/abs/1602.06559}{{\tt 1602.06559}}.

\bibitem{Bennett:2017kga}
E.~Bennett, D.~K. Hong, J.-W. Lee, C.~J.~D. Lin, B.~Lucini, M.~Piai et~al.,
  ``{Sp(4) gauge theory on the lattice: towards SU(4)/Sp(4) composite Higgs
  (and beyond)},''\href{http://dx.doi.org/10.1007/JHEP03(2018)185}{\emph{JHEP}
  {\bf 03} (2018) 185}, [\href{https://arxiv.org/abs/1712.04220}{{\tt
  1712.04220}}].

\bibitem{Lee:2018ztv}
J.-W. Lee, E.~Bennett, D.~K. Hong, C.~J.~D. Lin, B.~Lucini, M.~Piai et~al.,
  ``{Progress in the lattice simulations of Sp(2$N$) gauge
  theories},''\href{http://dx.doi.org/10.22323/1.334.0192}{\emph{PoS} {\bf
  LATTICE2018} (2018) 192}, [\href{https://arxiv.org/abs/1811.00276}{{\tt
  1811.00276}}].

\bibitem{Bennett:2019jzz}
Bennett, D.~K. Hong, J.-W. Lee, C.~J.~D. Lin, B.~Lucini, M.~Piai et~al.,
  ``{Sp(4) gauge theories on the lattice: $N_f=2$ dynamical fundamental
  fermions},'' \href{https://arxiv.org/abs/1909.12662}{{\tt 1909.12662}}.

\bibitem{Delbourgo:1982tv}
R.~Delbourgo and M.~D. Scadron, ``{Dynamical Symmetry Breaking and the Sigma
  Meson Mass in Quantum
  Chromodynamics},''\href{http://dx.doi.org/10.1103/PhysRevLett.48.379}{\emph{Phys.
  Rev. Lett.} {\bf 48} (1982) 379--382}.

\bibitem{Doff:2019vav}
A.~Doff and A.~A. Natale, ``{Technicolor models with coupled systems of
  Schwinger-Dyson
  equations},''\href{http://dx.doi.org/10.1103/PhysRevD.99.055026}{\emph{Phys.
  Rev.} {\bf D99} (2019) 055026}, [\href{https://arxiv.org/abs/1902.11072}{{\tt
  1902.11072}}].

\bibitem{Wang:2017tep}
L.~Wang, Z.~Fu and H.~Chen, ``{Hadronic coupling constants of
  $g_{\sigma\pi\pi}$ in lattice QCD},''
  \href{https://arxiv.org/abs/1702.08337}{{\tt 1702.08337}}.

\bibitem{Pelaez:2010fj}
J.~R. Pelaez and G.~Rios, ``{Chiral extrapolation of light resonances from one
  and two-loop unitarized Chiral Perturbation Theory versus lattice
  results},''\href{http://dx.doi.org/10.1103/PhysRevD.82.114002}{\emph{Phys.
  Rev.} {\bf D82} (2010) 114002}, [\href{https://arxiv.org/abs/1010.6008}{{\tt
  1010.6008}}].

\bibitem{Fichet:2016xpw}
S.~Fichet, G.~von Gersdorff, E.~Pont{\'o}n and R.~Rosenfeld, ``{The Global
  Higgs as a Signal for Compositeness at the
  LHC},''\href{http://dx.doi.org/10.1007/JHEP01(2017)012}{\emph{JHEP} {\bf 01}
  (2017) 012}, [\href{https://arxiv.org/abs/1608.01995}{{\tt 1608.01995}}].

\bibitem{Bando:1987br}
M.~Bando, T.~Kugo and K.~Yamawaki, ``{Nonlinear Realization and Hidden Local
  Symmetries},''\href{http://dx.doi.org/10.1016/0370-1573(88)90019-1}{\emph{Phys.
  Rept.} {\bf 164} (1988) 217--314}.

\end{thebibliography}\endgroup

\end{document}